\documentclass[conference, letterpaper]{IEEEtran}
\makeatletter
 \def\blfootnote{\xdef\@thefnmark{}\@footnotetext}
 \makeatother
\IEEEoverridecommandlockouts
\usepackage{cite}
\usepackage{amsmath,amssymb,amsfonts}
\usepackage{algorithmic}
\usepackage{algorithm}
\usepackage{caption}
\usepackage{graphicx}
\usepackage{textcomp}
\usepackage{xcolor}
\usepackage[inner=0.675in,outer=0.675in,top=0.673in,bottom=1.022in,pdftex]{geometry}
\usepackage{microtype}
\usepackage{subfigure}
\usepackage{booktabs} %
\usepackage{amsthm}
\usepackage{mdwmath}
\usepackage{mdwtab}
\usepackage{amsmath}
\usepackage{amsfonts}
\usepackage{enumitem}
\usepackage{dsfont}
\usepackage{hyperref}

\newtheorem{Definition}{Definition}
\newtheorem{Theorem}{Theorem}
\newtheorem{Lemma}{Lemma}
\newtheorem{Corollary}{Corollary}

\newtheorem{Remark}{Remark}

\DeclareMathAlphabet{\mathpzc}{OT1}{pzc}{m}{it}
\begin{document}

\title{On the Privacy Guarantees of Gossip Protocols in General Networks}

\author{
    \IEEEauthorblockN{Richeng Jin, \textit{Student Member, IEEE}, Yufan Huang, \textit{Student Member, IEEE}, \\
    Huaiyu Dai, \textit{Fellow, IEEE}}
}
\maketitle
\begin{abstract}
Recently, the privacy guarantees of information dissemination protocols have attracted increasing research interests, among which the gossip protocols assume vital importance in various information exchange applications. In this work, we study the privacy guarantees of gossip protocols in general networks in terms of differential privacy and prediction uncertainty. First, lower bounds of the differential privacy guarantees are derived for gossip protocols in general networks in both synchronous and asynchronous settings. The prediction uncertainty of the source node given a uniform prior is also determined. For the private gossip algorithm, the differential privacy and prediction uncertainty guarantees are derived in closed form. Moreover, considering that these two metrics may be restrictive in some scenarios, the relaxed variants are proposed. It is found that source anonymity is closely related to some key network structure parameters in the general network setting. Then, we investigate information spreading in wireless networks with unreliable communications, and quantify the tradeoff between differential privacy guarantees and information spreading efficiency. Finally, considering that the attacker may not be present at the beginning of the information dissemination process, the scenario of delayed monitoring is studied and the corresponding differential privacy guarantees are evaluated.
\end{abstract}
{\blfootnote{R. Jin, Y. Huang and H. Dai are with the Department of Electrical and Computer Engineering, North Carolina State University, Raleigh, NC, USA, 27695 (e-mail: {rjin2,yhuang20,hdai}@ncsu.edu).
}}

\section{Introduction}
\noindent It is well-known that most people are six or fewer social connections away from each other. Recently, the explosive development in the Internet and social networks makes it easy for people to disseminate their information to the rest of the world. Gossip protocols, in which networked nodes randomly choose a neighbor to exchange information, have been widely adopted in various applications for information dissemination due to their simplicity and efficiency. For instance, they can be used to spread and aggregate information in dynamic networks like mobile networks, wireless sensor networks, and unstructured P2P networks \cite{chandra2001anonymous,dimakis2008geographic,ganesh2003peer}. Combined with stochastic gradient descent methods, gossip protocols are also adapted to implement distributed machine learning \cite{bianchi2013convergence,liu2018differentially}. In particular, the authors of \cite{liu2018differentially} propose to transmit differentially private gradient information through gossip protocols. Nonetheless, they focus on the privacy of the shared gradient information rather than the anonymity of the source.

With the arising concerns of privacy exposure, the information sources often prefer to stay anonymous while disseminating some sensitive information. Gossip protocols are believed to provide a certain form of source anonymity since most nodes don't get informed directly from the source, and the origin of the information becomes increasing blurred as the spreading proceeds. In this regard, source identification and protection of gossip protocols have attracted significant research interests (see \cite{jiang2016identifying,fanti2017hiding} and the references therein). However, the existing approaches usually assume some specific network structures (e.g., tree graphs) and attacking techniques (e.g., maximum likelihood estimator) and don't easily generalize.

To study the privacy of gossip protocols in a formal and rigorous setting, the concept of differential privacy \cite{dwork2014algorithmic}, which was originally introduced in data science, is adapted to measure the source anonymity of gossip protocols in \cite{bellet2019started}. However, their study is restricted to complete networks, which may not be a good model in practice. For example, practical networks often have a network diameter much larger than $1$ ($41$ for the Facebook network \cite{backstrom2012four}).

In this work, we extend the study of the fundamental limits on the privacy of gossip-based information spreading protocols to general networks. Our main contributions are summarized as follows.
\begin{enumerate}
\item Lower bounds of the differential privacy guarantees of general gossip protocols are derived for general networks in both synchronous and asynchronous settings. The prediction uncertainty of the source node given a uniform prior is also determined.
\item For the private gossip algorithm \cite{bellet2019started}, the differential privacy guarantees and prediction uncertainty are derived in closed form. In addition, considering that the original differential privacy and prediction uncertainty may be restrictive in some scenarios, candidate set based differential privacy and prediction uncertainty variants are proposed. Numerical results are presented to show the privacy guarantees of the private gossip algorithm in different network structures.
\item The privacy guarantees of standard gossip and private gossip protocols are further studied in a wireless setting, where communications are assumed to be unreliable. It is found that wireless interference can enhance the differential privacy while slowing down the spreading process. Through analysis and simulations, the tradeoff between the differential privacy guarantees and the information spreading efficiency is revealed.
\item The effect of the additional uncertainty induced by delayed monitoring on the privacy guarantees is shown.
\end{enumerate}

The remainder of this work is organized as follows. Section~\ref{PF} introduces the System model. The privacy guarantee of gossip protocols in general networks are presented in Section \ref{Privacy_of_Gossip_Protocols}. The privacy-spreading tradeoff of gossip protocols in wireless networks is discussed in Section \ref{tradeoffprivacyspreading}. Section \ref{delatedminitoringprivacy} investigates the privacy guarantees of gossip protocols in the delayed monitoring scenario. Conclusion and future works are discussed in Section \ref{CF}.

\section{System Model}\label{PF}
\subsection{Gossip Protocol}
\noindent In this work, we investigate the privacy of information source in gossip-based information spreading. The goal is to measure the capability of gossip protocols in keeping the information source anonymous. Specifically, given a connected network $G=(V,E)$ of arbitrary topology, where $V=\{0,1,...,n-1\}$ is the node set and $E$ is the set of connecting edges, a node (source) initially possesses a piece of information and needs to deliver it to all the other nodes in the network. All the nodes are assumed to share the same communication protocol $gossip$. Each time an informed node $i$ performs $gossip$, it will contact one of its neighboring nodes $j\in N_i$ uniformly at random (i.e., with probability $1/d_i$, where $d_i$ is the degree of node $i$). The whole information dissemination process terminates after all the nodes are informed. Same as \cite{bellet2019started}, we focus on the gossip protocols based on the ``push'' action\footnote{In the corresponding ``pull" action, uninformed nodes are active and try to solicit the information from informed nodes. The ``push'' action is dominant for information spreading in social and mobile networks. In addition, such a study is also conservative in the sense that it gives the attacker an advantage by only monitoring the ``push'' actions.} in this work, and consider the following two specific gossip protocols.
\begin{enumerate}
\item Standard Gossip: All informed nodes remain active (i.e., continuously performing $gossip$) during the spreading process.
\item Private Gossip \cite{bellet2019started} (i.e., Algorithm \ref{PrivateGossip}): Once an active informed node (initially it is the source) performs \textit{gossip}, it turns inactive, and the newly informed node takes over the source role.
\end{enumerate}

\begin{algorithm}
\caption{Private Gossip Algorithm \cite{bellet2019started}}
\label{PrivateGossip}
\begin{algorithmic}[1]
\STATE \textbf{Require:} The number of nodes $n$, the source node $k$.
\STATE \textbf{Ensure:} The informed node set $I=\{0,1,\cdots,n-1\}$
\STATE \textbf{Initialization:} Informed node set $I \leftarrow \{k\}$, active node set $A^{c} \leftarrow \{k\}$
\WHILE{$|I| < n$}{
\STATE The active node in $A^{c}$ performs $gossip$ and another node $j$ is informed
\STATE $I \leftarrow I \cup \{j\}, A^{c} \leftarrow \{j\}$}
\ENDWHILE
\end{algorithmic}
\end{algorithm}

\subsection{Time Model}
\noindent Both synchronous and asynchronous time models are considered. In the former, all nodes share a global discrete time clock. Each time the clock ticks, all active informed nodes perform the $gossip$ action simultaneously, and the informed node set is updated accordingly, counted as one round. In the asynchronous time model, each node has its own internal clock, which ticks according to a Poisson process, with the mean interval between two ticks equivalent to that of one round in the synchronous model. The $gossip$ action and update of the informed node set is performed each time the clock of an active informed node ticks.

\subsection{Threat Model}\label{observationmodel}
\begin{figure}[t]
  \centering
  \centerline{\includegraphics[width=3.5in]{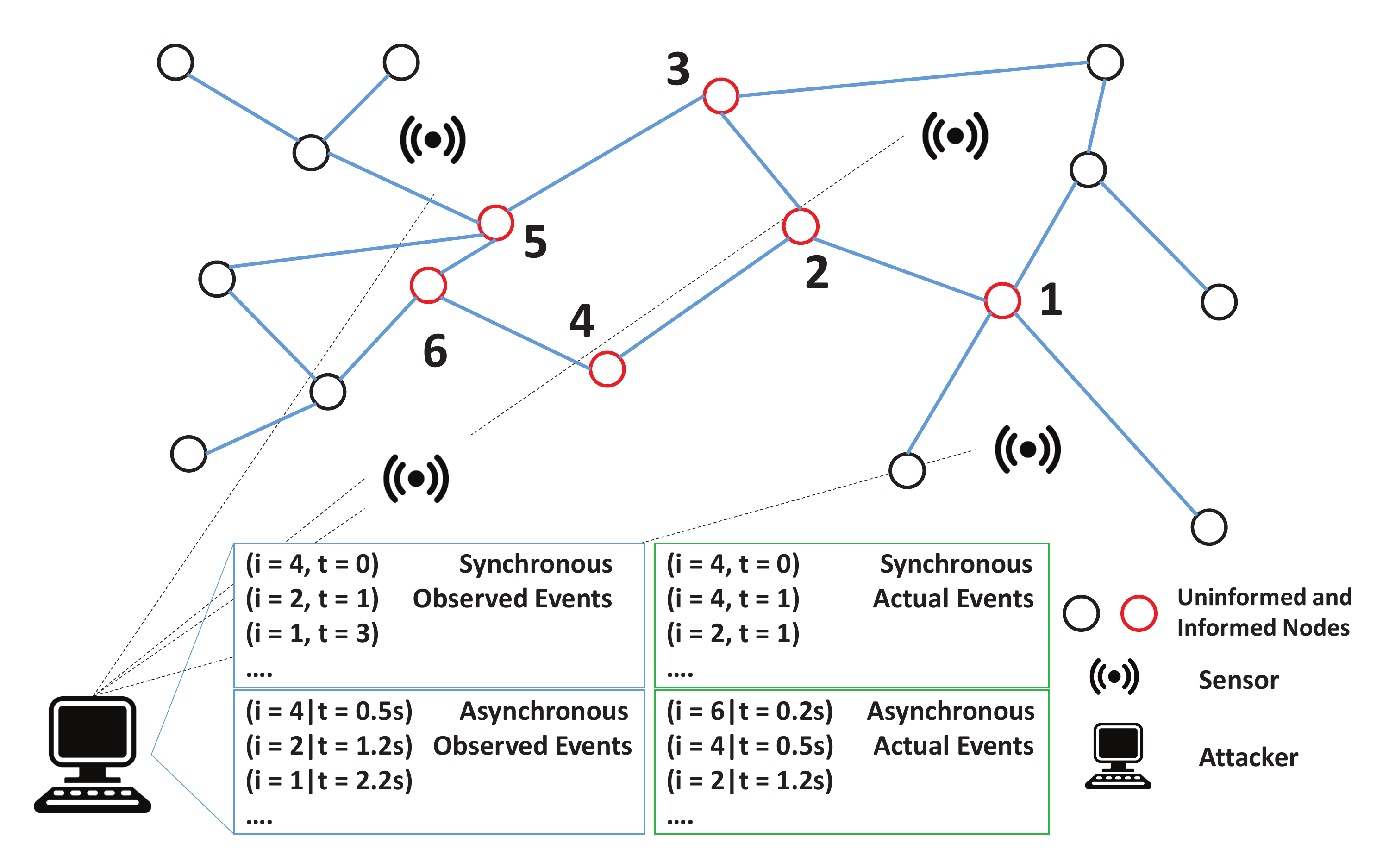}}
\caption{Sensor Monitoring and Observations}\label{F0}
\vspace{-0.2in}
\end{figure}

\noindent The goal of the attacker is to identify the source node based on its observations (i.e., attack on confidentiality and privacy). It is assumed that the attacker can monitor the ongoing communications in the whole network, through, e.g., deploying a sufficient number of sensors throughout the field. With a probability of $0 < \alpha \leq 1$, the sensors can correctly observe the identities of the active nodes at each gossip step. Specifically, as shown in Fig. \ref{F0}, the observed event has the form of $\mathpzc{S}=((i,t)),\ i\in V,\ t\in \{0,1,2,\cdots\}$ in the synchronous setting, which indicates that the attacker knows node $i$ performs the gossip action at time slot $t$. In the asynchronous setting, however, the attacker does not know the exact time of each observed event, but only the relative order of the nodes' activities. The observed event in this case is represented by $\mathpzc{S}=((i|t))$, where the condition $t$ stands for the latent time information unknown to the attacker.

\subsection{Privacy Model}
\noindent In this work, differential privacy is adopted to measure the information leakage of the gossip protocols. In particular, a randomized algorithm $\mathpzc{R}$ with domain $\mathbb{N}^{|\chi|}$ is $(\epsilon, \delta)$-differentially private if for all $\mathpzc{S}\subseteq Range(\mathpzc{R})$ and for any two databases $x,y$ that differ on a single element \cite{dwork2014algorithmic}:
\begin{equation}\label{EqOD}
Pr[\mathpzc{R}(x)\in \mathpzc{S}]\leq e^{\epsilon}Pr[\mathpzc{R}(y)\in \mathpzc{S}] +\delta,
\end{equation}
where parameter $\epsilon \geq 0$ is the privacy budget while $\delta \geq 0$ is the tolerance level for the violation of the $\epsilon$ bound. Specifically, given the privacy budget $\epsilon$ and the tolerance level $\delta$, Eq. (\ref{EqOD}) implies that the randomized algorithm guarantees that the privacy loss is bounded by $\epsilon$ with a probability of at least $1-\delta$. Intuitively, the smaller the $\epsilon$ and the $\delta$, the better the privacy. Consider a source indicator database of the format $\mathpzc{D}^{(i)}=[0,...,\mathpzc{d}_i=1,...,0]$ with exactly one nonzero value $\mathpzc{d}_i=1$ if node $i$ is the source. Given $D\triangleq\{\mathpzc{D}^{(i)}\}_{i=0}^{n-1}$ and the graph $G$ as the input, a gossip protocol can be treated as a randomized algorithm with the output set $\mathbb{S}$ (i.e., the range) consisting of all possible observation sequences by the attacker during the execution of the protocol.
\begin{Definition}\label{definition1}
Given a general network $G$, a gossip protocol is $(\epsilon, \delta)$-differentially private in $G$ if for all observations $\mathpzc{S}\subseteq \mathbb{S}$ and for any two source indicator vectors $\mathpzc{D}^{(i)}, \mathpzc{D}^{(j)}, i,j\in V$:
\begin{equation}\label{EqM}
p_G^{(i)}(\mathpzc{S})\leq e^{\epsilon}p_G^{(j)}(\mathpzc{S})+\delta,
\end{equation}
where $p_G^{(i)}(\mathpzc{S})=Pr[\mathpzc{S}|G,\mathpzc{D}^{(i)}]$ is the conditional probability of an observation event $\mathpzc{S}$ given the network graph $G$ and the source indicator vector $D^{(i)}$.
\end{Definition}

In this work, considering the fact that, due to the topological and observation model constraints, there may exist some (rare) events $\mathpzc{S}$ such that $p_G^{(j)}(\mathpzc{S})=0$ (e.g., if $\mathpzc{S}_{i,0}$ is the observed event that node $i$ performs \textit{gossip} at time $0$ in the synchronous setting, then $p_G^{(j)}(\mathpzc{S}_{i,0})=0, \forall j\neq i$), additional tolerance level is needed to ensure the privacy guarantees. Thus, for the privacy guarantees of general gossip protocols, we mainly focus on the study of the tolerance level $\delta$. For the private gossip algorithm (i.e., Algorithm \ref{PrivateGossip}) in the asynchronous setting, we are able to derive the corresponding differential privacy level $\epsilon$ and give some analysis.


In addition to differential privacy, it is also desirable to study privacy guarantees of information dissemination protocols from a more pertinent perspective, i.e., source identification through prediction or detection. Reusing the above example, there always exist some events $\mathpzc{S}$ such that $p_G^{(i)}(\mathpzc{S}) > 0$ for some $i$ but $p_G^{(j)}(\mathpzc{S}) = 0, \forall j \neq i \in V$, which satisfy an arbitrary privacy budget $\epsilon$ with a tolerance level of $\delta$ (if $p_G^{(i)}(\mathpzc{S}) \leq \delta$). However, the identity of the source (i.e., node $i$) can still be easily inferred. Therefore, it is further required that some \textit{prediction uncertainty} be guaranteed for a given differentially private protocol, which is defined as \cite{bellet2019started}:
\begin{Definition}
Given a general network $G$, the prediction uncertainty of a gossip protocol is defined for a uniform prior $p_G(I_0)$ on source nodes and any $i\in \{0,1,...,n-1\}$ as:
\begin{equation}
\begin{split}
c &= \min_{i,\mathpzc{S}\subseteq \mathbb{S}}\left(\frac{p_G(I_0\neq\{i\}|\mathpzc{S})}{p_G(I_0=\{i\}|\mathpzc{S})}\right)\\
&= \min_{i,\mathpzc{S}\subseteq \mathbb{S}}\left(\frac{1}{p_G(I_0=\{i\}|\mathpzc{S})}\right)-1,~ \forall p_G^{(i)}(\mathpzc{S})>0,
\end{split}
\end{equation}
\end{Definition}
where $I_0$ stands for the initial informed node set and its element represents the source node.
\begin{Remark}
The connection of prediction uncertainty and differential privacy is illustrated below. If the attacker obtains an observation $\mathpzc{S}$, differential privacy measures the probabilities of it observing $\mathpzc{S}$ given different sources while prediction uncertainty considers the posterior probabilities of different sources given $\mathpzc{S}$. Especially, because of the uniform prior $p_G(I_0)$, $\frac{p_G(I_0\neq\{i\}|\mathpzc{S})}{p_G(I_0=\{i\}|\mathpzc{S})}=\frac{\sum_{j\neq i}p_G^{(j)}(\mathpzc{S})}{p_G^{(i)}(\mathpzc{S})}$ holds by the Bayes' formula. Prediction uncertainty is an appealing metric in this study as it measures the privacy guarantees from the source prediction perspective with a much smaller cardinality than the classic privacy budget (which requires the study of all pairs of $p_G^{(i)}(\mathpzc{S})$ and $p_G^{(j)}(\mathpzc{S})$). Moreover, given a prediction uncertainty $c$, it can be shown that $p_G(I_0=\{i\}|\mathpzc{S})\leq \frac{1}{c+1},\forall i,\mathpzc{S}$; therefore a larger $c$ indicates better source anonymity.
\end{Remark}

In order to further remedy the aforementioned issue of differential privacy, a relaxed differential privacy variant, termed differential privacy within a candidate set, is proposed. Its definition is given as follows.

\begin{Definition}
Given a general network $G$, a gossip protocol is $(\epsilon, \delta)$-differentially private within a candidate set $\mathpzc{Q}$ in $G$ if for all observations $\mathpzc{S}\subseteq \mathbb{S}$ and for any two source indicator vectors $\mathpzc{D}^{(i)}, \mathpzc{D}^{(j)}, i,j\in \mathpzc{Q} \subseteq V$:
\begin{equation}
p_G^{(i)}(\mathpzc{S})\leq e^{\epsilon}p_G^{(j)}(\mathpzc{S})+\delta,
\end{equation}
where $p_G^{(i)}(\mathpzc{S})=Pr[\mathpzc{S}|G,\mathpzc{D}^{(i)}]$ is the conditional probability of an observation event $\mathpzc{S}$ given the network graph $G$ and the source indicator vector $D^{(i)}$.
\end{Definition}

\begin{Remark}
The only difference between differential privacy and differential privacy within a candidate set is that the privacy guarantee is ensured for the source nodes falling in a candidate set $\mathpzc{Q}$ instead of the whole network $V$. Note that the notion of differential privacy is highly conservative (considering the worst-case scenario). If there exists a node $i$ such that $p_{G}^{(i)}(\mathpzc{S}) = 0$, the probability of node $i$ being the source is 0 from the attacker's perspective. In such a case, the attacker can always differentiate node $i$ from other nodes, which means that the differential privacy guarantee will be violated with a high probability. However, on the one hand, it does not necessarily mean that the gossip protocols cannot provide effective source privacy protection. Particularly, it may be difficult for the attacker to distinguish any pairs of other nodes except node $i$ in the network (e.g., $p_{G}^{j}(\mathpzc{S}) = p_{G}^{z}(\mathpzc{S}), \forall j\neq z \in V\setminus \{i\}$). On the other hand, in practice, it may not be necessary for the source node to hide itself in the whole network. Instead, it may be enough to make itself indistinguishable among a subset of the network (e.g., its neighbors). Our definition of differential privacy within a candidate set measures the privacy guarantee of the gossip protocols in such scenarios. Especially, as long as $p_{G}^{(j)}(\mathpzc{S}) > 0, \forall j\in \mathpzc{Q}$, $\delta = 0$ is always feasible for differential privacy within the candidate set $\mathpzc{Q}$. In addition, when $\mathpzc{Q} = V$, it is equivalent to the original definition of differential privacy. Therefore, it can be understood as a relaxed version of differential privacy.
\end{Remark}

Similarly, the prediction uncertainty within a candidate set is defined as follows.

\begin{Definition}
Given a general network $G$, the prediction uncertainty of a gossip protocol within the candidate set $\mathpzc{Q}$ is defined for a uniform prior $p_G(I_0)$ on source nodes over the candidate set $\mathpzc{Q}$ as:
\begin{equation}
c = \min_{i\in \mathpzc{Q},\mathpzc{S}\subseteq \mathbb{S}}\left(\frac{\sum_{j\neq i \in \mathpzc{Q}}p_G(I_0 = \{j\}|\mathpzc{S})}{p_G(I_0=\{i\}|\mathpzc{S})}\right),\ \forall p_G^{(i)}(\mathpzc{S})>0.
\end{equation}
\end{Definition}

\section{Privacy of Gossip Protocols in General Networks}\label{Privacy_of_Gossip_Protocols}
In this section, the privacy guarantees of gossip protocols are investigated. For the general gossip protocols, general results concerning the lower bounds of the tolerance level $\delta$ and the upper bounds of the prediction uncertainty $c$ are obtained. For the private gossip algorithm in the asynchronous setting, the pure version of differential privacy ($\delta = 0$) is feasible, and the corresponding differential privacy level $\epsilon$ and prediction uncertainty $c$ are derived.
\subsection{General Gossip Protocols}
\noindent In this subsection, the privacy guarantees of general gossip protocols in general networks are studied. To facilitate our following analysis, we need the following lemma and the definition of \textit{decay centrality}.
\begin{Lemma}\label{Lemma1}
Given any gossip protocol in a graph $G$, let $\mathpzc{S}\subseteq\mathbb{S}$ and there are two constants $w_G^{(i)}(\mathpzc{S}),w_G^{(j)}(\mathpzc{S})$ such that $p_G^{(i)}(\mathpzc{S})\geq w_G^{(i)}(\mathpzc{S})$ and $p_G^{(j)}(\mathpzc{S})\leq w_G^{(j)}(\mathpzc{S})$. If the gossip protocol satisfies $(\epsilon,\delta)$-differential privacy, then $\delta\geq \max_{\mathpzc{S},i,j}(w_G^{(i)}(\mathpzc{S})-e^{\epsilon}w_G^{(j)}(\mathpzc{S}))$.
\end{Lemma}

Lemma 1 readily follows from the definition of differential privacy; its proof is omitted in the interest of space.

\begin{Definition}
\cite{jackson2010social} Given a network $G$ and a decay parameter $\beta$, $0<\beta<1$, the \textbf{decay centrality} of node $i$ is defined as
\begin{equation}
C_\beta(i)=\sum_{j\neq i}\beta^{d(i,j)},
\end{equation}
where $d(i,j)$ is the length of the shortest path between node $i$ and $j$.
\end{Definition}

\begin{Remark}
Decay centrality measures the ease of a node reaching out to other nodes in the network. A large decay centrality indicates the central positioning of a node and its easiness to reach other nodes. The difficulty increases as $\beta$ decreases.
\end{Remark}

Our main result concerning the privacy guarantees of general gossip protocols in a general network is given below.

\begin{Theorem}\label{T1}
Given a connected network $G$ with $n$ nodes and diameter $D_G=\max_{i,j\in V, i\neq j}d(i,j)$, and considering the observation model described in Section \ref{observationmodel} with parameter $\alpha$, if a gossip protocol satisfies $(\epsilon, \delta)$-differential privacy for any $\epsilon \geq 0$ and $c$-prediction uncertainty, then we have $\delta\geq \alpha$ and $c=0$ in the synchronous setting. In the asynchronous setting,
\begin{equation}\label{EqPR}
\begin{aligned}
\delta&\geq \max\left[\alpha-e^{\epsilon}(1-\alpha)^{D_G}, \alpha-e^{\epsilon}\frac{1-\alpha}{n-1}\right]
\end{aligned}
\end{equation}
and
\begin{equation}\label{EqPU}
c\leq \min_{i\in V} \frac{C_{1-\alpha}(i)}{\alpha},
\end{equation}
where $C_{1-\alpha}(i)$ is the decay centrality of node $i$ with decay parameter $1-\alpha$.
\end{Theorem}
\begin{IEEEproof}
Please see Appendix \ref{ProofTheoremT1}.
\end{IEEEproof}

\begin{Remark}
Some interpretations of the results of Theorem 1 are in order. It can be observed that the asynchronous setting provides better privacy guarantees than the synchronous setting, since the attacker has less information (i.e., the timing of the events) in this case. Note that differential privacy considers the worst case scenario. In the synchronous setting, when the attacker detects the activity of a node at time $0$, it can infer that the corresponding node is the source immediately. Therefore, the prediction uncertainty is $0$ due to this worst-case event, and the privacy guarantees are determined by the attacker's sensing capability $\alpha$ in the synchronous setting. In the asynchronous setting, however, the attacker could not directly infer the source solely based on the first-observed event due to the lack of associated timing. A counter example can be found in Fig. \ref{F0}.

\begin{figure}
\centering
\begin{minipage}{.24\textwidth}
  \centering
  \captionsetup{justification=centering}
  \includegraphics[width=\linewidth]{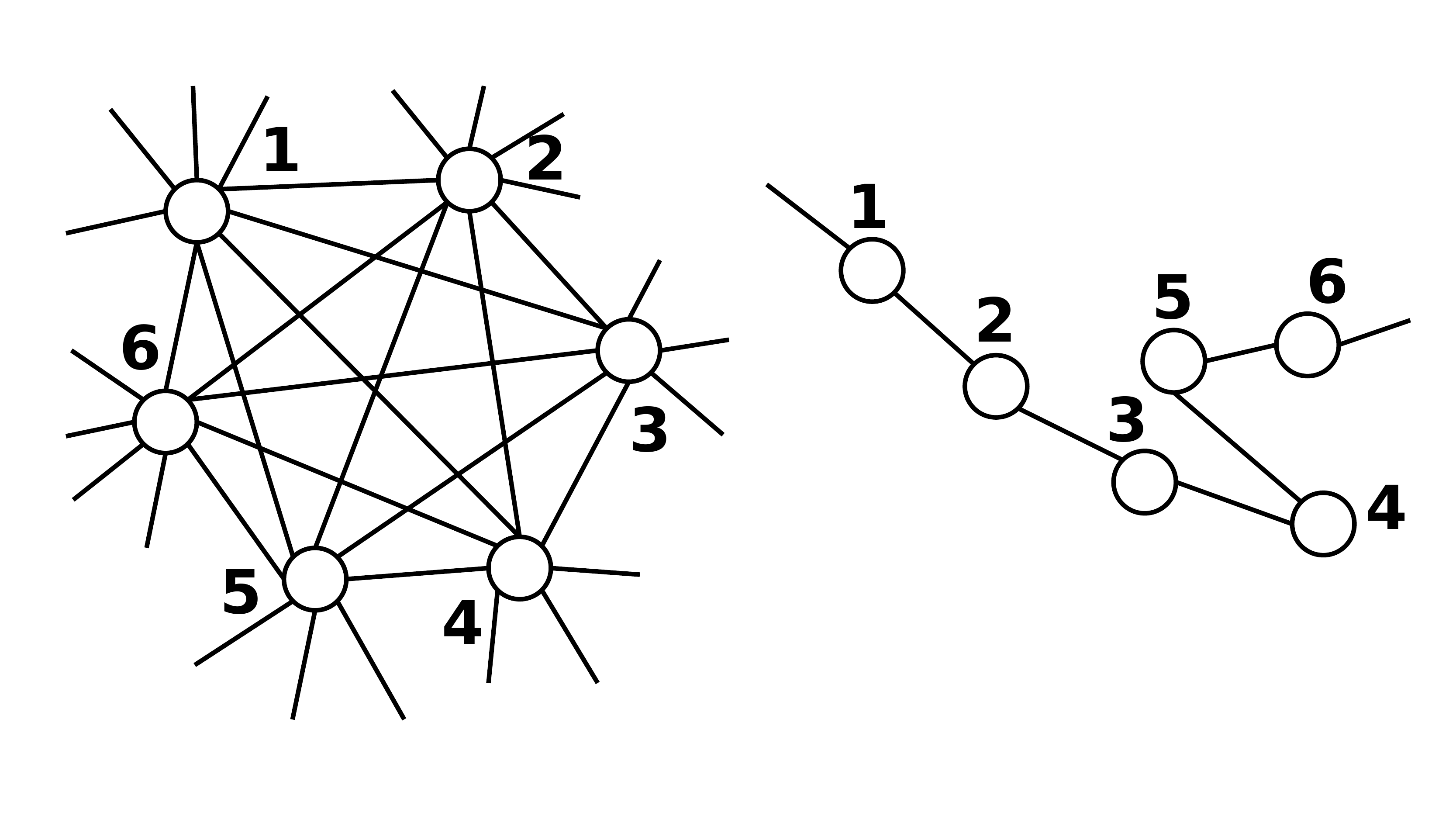}
  \caption{Node 1 and node 6 are more distinguishable in the right network.}
  \label{diameter}
\end{minipage}%
\begin{minipage}{.24\textwidth}
  \centering
  \captionsetup{justification=centering}
  \includegraphics[width=\linewidth]{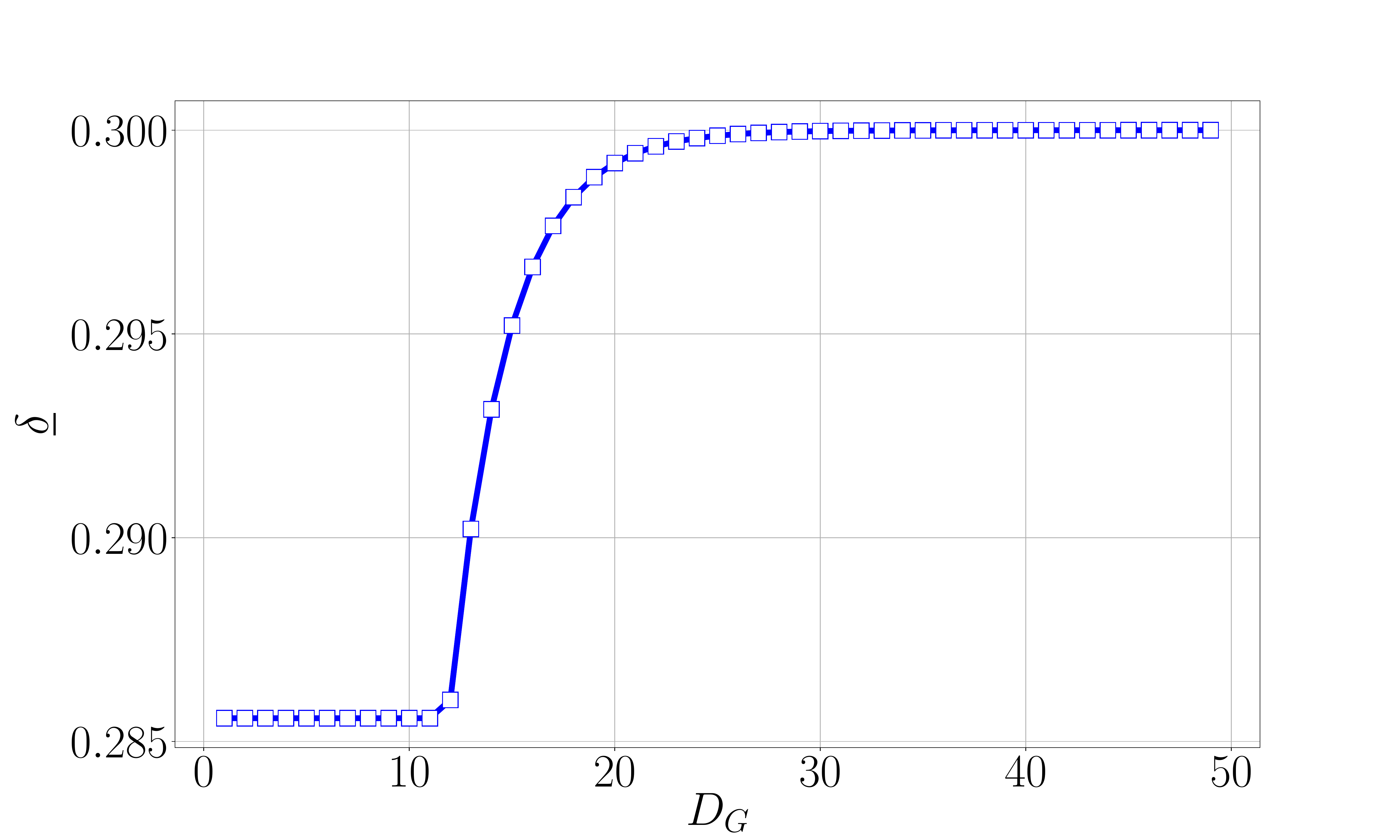}
  \caption{Tolerance Level v.s. Network Diameter ($\alpha=0.3, n=50, \epsilon=0.01$).}
  \label{PvD}
\end{minipage}
\vspace{-0.2in}
\end{figure}

As a result, the structure of the network plays an important role in the asynchronous setting. In the context of information spreading, if two nodes are further apart, it takes more time for the information to be spread from one to the other; this duration gives the attacker more opportunities to differentiate the detected events, which leads to potentially higher privacy loss of the source node’s identity. For instance, in the left network of Fig. \ref{diameter}, considering the event $\mathpzc{S}_{1,0}$, i.e., node $1$'s activity being the first observed event by the attacker in the asynchronous setting, the probability of this event given that the source is $6$ is $p_G^{(6)}(\mathpzc{S}_{1,0})\leq (1-\alpha)$ according to (\ref{EqFP}) and the probability of this event given that the source is $1$ is $p_G^{(1)}(\mathpzc{S}_{1,0}) \geq \alpha$. But in the right network, the corresponding probabilities are $p_G^{(6)}(\mathpzc{S}_{1,0})\leq (1-\alpha)^5$ and $p_G^{(1)}(\mathpzc{S}_{1,0}) \geq \alpha$, which makes $\mathpzc{S}_{1,0}$ a more distinguishable event in the right network. Therefore, the network diameter $D_G$, as the distance measure of the whole network, captures the potential privacy loss and becomes a key factor of the differential privacy lower bound in (\ref{EqPR}); an example of the relationship between the differential privacy tolerance level bound and the network diameter is shown in Fig. \ref{PvD}. The same logic is reflected on the prediction uncertainty given in (\ref{EqPU}). The smaller the decay centrality a network has (i.e., the nodes are more distant from each other), the more likely the attacker can identify the source node through its observations. Therefore, the inherent network structure imposes certain limit on privacy preserving concerning the source node identity, which applies to all information spreading protocols and calls for other privacy protection mechanisms, to be further explored in future work.

In addition, it can be seen that as the attacker's sensing capability $\alpha$ increases the privacy guarantees decrease (i.e., $\delta$ increases and $c$ decreases). In particular, for an omnipresent attacker with $\alpha = 1$, we have $\delta=1$ and $c=0$ even in the asynchronous setting.

Finally, given (\ref{EqPR}), the lower bound of the differential privacy level $\epsilon$ in the asynchronous setting for a given $\delta$ can be obtained as follows
\begin{equation}\label{epsilonlowerbound}
e^\epsilon \geq \max\left[\left(\frac{\alpha - \delta}{(1-\alpha)^{D_{G}}}\right), \left(\frac{(\alpha - \delta)(n-1)}{1-\alpha}\right)\right].
\end{equation}

\end{Remark}

\subsection{Private Gossip Algorithm}
\noindent In this subsection, the differential privacy level $\epsilon$ (with the tolerance level $\delta = 0$) and the prediction uncertainty of Algorithm \ref{PrivateGossip} in the asynchronous setting is examined.\footnote{Note that in the synchronous setting, $\delta = 0$ may not be achievable.}

\begin{figure}[t]
  \centering
  \centerline{\includegraphics[width=2in]{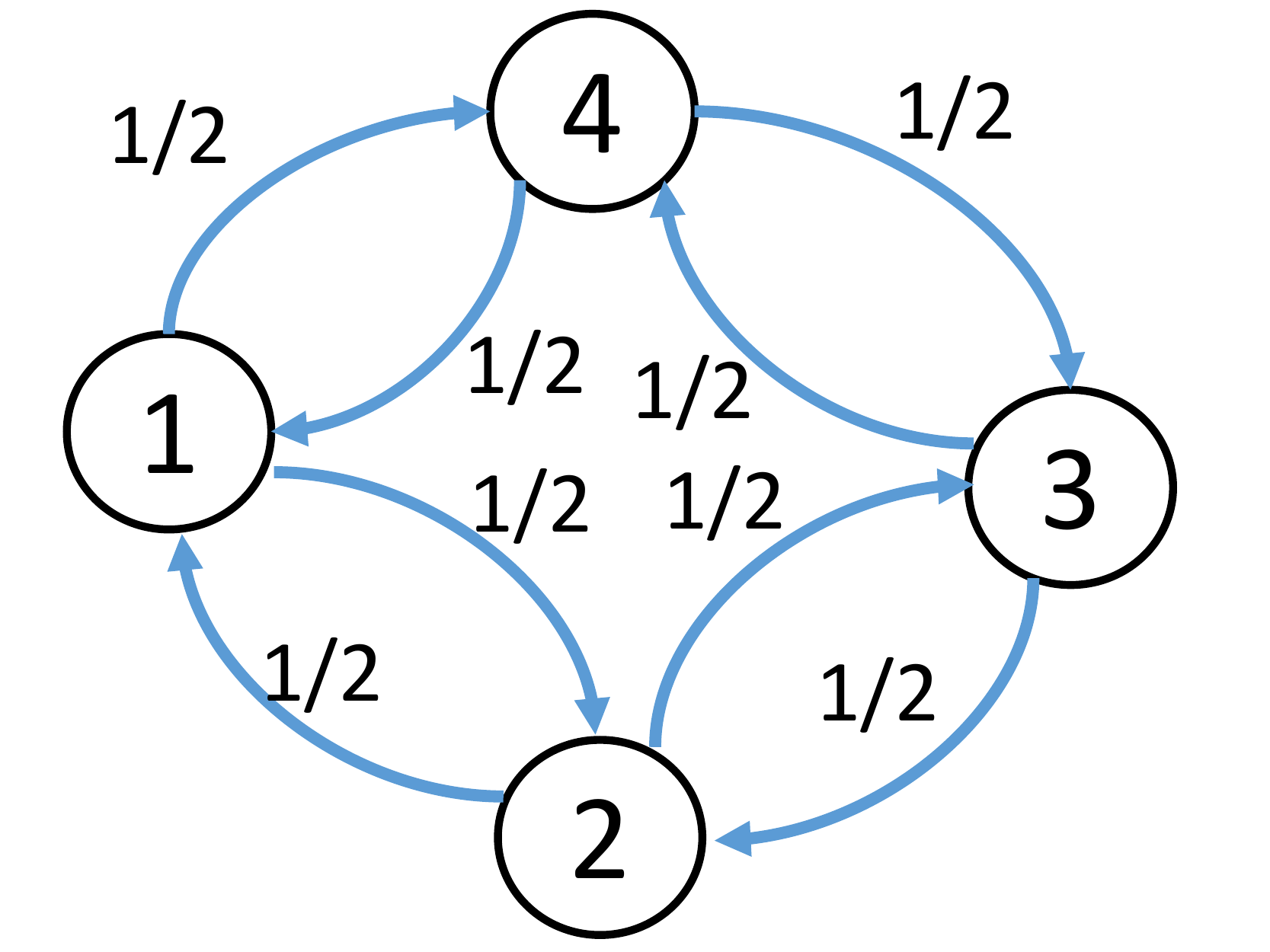}}
\caption{The index of the active node in Algorithm \ref{PrivateGossip} for a ring graph with 4 nodes. Each node has two neighbors and the probability of it contacting each neighboring node is $\frac{1}{2}$. For instance, if node $1$ is active (i.e., the Markov chain is in state $1$) at time $m-1$, then the probabilities of node $2$ and node $4$ being active at time $m$ are both $\frac{1}{2}$.}\label{MarkovChain}
\end{figure}

In this case, since only one node is active at each time slot, the index of the active node can be modeled as a Markov chain with a transition probability matrix $A$. The $(j,i)$-th entry of $A$, denoted by $A[j,i]$, measures the probability of node $j$ contacting node $i$ once it becomes active. Fig. \ref{MarkovChain} shows an exemplary Markov chain of a ring graph with 4 nodes. In addition, to facilitate the discussion, we further define another matrix $\hat{A}_{i}$ as follows.

\begin{equation}\label{modifiedmatrix}
\hat{A}_{i}[j,k] =
\begin{cases}
A[j,k], \hfill \text{if $j \neq i$},\\
0, \hfill \text{~~~~~~~~~~~if $j = i$ and $j \neq k$},\\
1, \hfill \text{if $j=k=i$}.
\end{cases}
\end{equation}

\begin{Remark}\label{remark5}
Note that $\hat{A}_{i}$ corresponds to the transition probability matrix of the Markov chain by setting node $i$ as an absorbing state (i.e., node $i$ will not push its message to any other nodes). Let $\hat{A}_{i}^{m}$ denote the $m$-th power of the matrix $\hat{A}_{i}$. Then, given the source node $j$, $\hat{A}_{i}^{m}[j,i]$ and $\hat{A}_{i}^{m-1}[j,i]$ are the probabilities of node $i$ being active at time $m$ and $m-1$, respectively. Note that if node $i$ is active at time $m-1$, it is active at time $m$ with probability 1 since it is an absorbing state. In this sense, given the source node $j$, the probability of node $i$ being active for the first time at time $m$ is $\hat{A}_{i}^{m}[j,i] - \hat{A}_{i}^{m-1}[j,i]$.
\end{Remark}

To facilitate the discussion, we introduce the following quantity.

\begin{Definition}
The probability of \textbf{secret message spreading} from the source node $j$ to another node $i$, denoted by $P(j \rightarrow i)$, is defined as the probability that, given the source node $j$, node $i$ becomes active before the attacker observes the first event.
\end{Definition}

\begin{Remark}
In the asynchronous setting, $P(j \rightarrow i)$ measures the similarity between node $j$ and node $i$ from the attacker's perspective. Intuitively, given source node $j$, if node $i$ becomes active before the attacker observes the first event, the attacker cannot differentiate node $j$ and node $i$ based on its observation. The larger the $P(j \rightarrow i)$, more difficult it is for the attacker to differentiate node $j$ and node $i$.
\end{Remark}

With such consideration, the following lemma can be proved.

\begin{Lemma}\label{TheoremProbability}
For Algorithm \ref{PrivateGossip}, given a general network $G$ with source node $j$ and the observation model described in Section \ref{observationmodel} with parameter $\alpha$, the probability of secret message spreading from the source node $j$ to another node $i$ for the first time is given by
\begin{equation}
P(j \rightarrow i) = \alpha \sum_{m=0}^{\infty}(1-\alpha)^{m}\hat{A}_{i}^{m}[j,i] = \alpha(I-(1-\alpha)\hat{A}_{i})^{-1}[j,i],
\end{equation}
where $I$ is an identity matrix and $\hat{A}_{i}$ is the transition probability matrix defined in (\ref{modifiedmatrix}).
\end{Lemma}
\begin{IEEEproof}
Please see Appendix \ref{ProofTheoremProbability}.
\end{IEEEproof}

Given Lemma \ref{TheoremProbability}, the privacy guarantees of Algorithm \ref{PrivateGossip} can be shown as follows.
\begin{Theorem}\label{dpprivategossip}
Given a general network $G$ and the observation model described in Section \ref{observationmodel} with parameter $\alpha$, Algorithm \ref{PrivateGossip} is $(\epsilon,0)$-differentially private in the asynchronous setting, where
\begin{equation}\label{epsilonstandard}
\epsilon = \ln\bigg(\max_{j \neq i \in V}\frac{1}{\alpha(I-(1-\alpha)\hat{A}_{i})^{-1}[j,i]}\bigg).
\end{equation}
The prediction uncertainty of Algorithm \ref{PrivateGossip} in the asynchronous setting is given by
\begin{equation}
c = \min_{i\in V}\sum_{j \neq i\in V}\alpha(I-(1-\alpha)\hat{A}_{i}))^{-1}[j,i].
\end{equation}
\end{Theorem}
\begin{IEEEproof}
Please see Appendix \ref{Proofdpprivategossip}.
\end{IEEEproof}

\begin{Remark}
Note that $\epsilon = \ln\left(\max_{j\neq i \in V}\frac{1}{P(j\rightarrow i)}\right)$ and $P(j \rightarrow i)=\alpha \sum_{m=0}^{\infty}(1-\alpha)^{m}\hat{A}_{i}^{m}[j,i]$. For any $m < d(j,i)$, it can be verified that $\hat{A}_{i}^{m}[j,i] = 0$. On the other hand, $(1-\alpha)^{m}$ decreases exponentially. In this sense, $P(j \rightarrow i)$ is supposed to decrease (and therefore $\epsilon$ will increase) as $d(j,i)$ increases, which verifies our discussion about the importance of the network structure (i.e., diameter) on differential privacy.
\end{Remark}
In particular, given the $(\epsilon,0)$-differential privacy guarantee, the corresponding tolerance level $\delta(\epsilon')$ can be obtained for any $\epsilon' > 0$ using the following corollary.

\begin{Corollary}\label{corollarydp}
Any $(\epsilon,0)$-differentially private mechanism is also $(\epsilon',\delta(\epsilon'))$-differentially private for all $\epsilon'>0$, where
\begin{equation}
\delta(\epsilon') = \Phi\left(-\frac{\epsilon'}{\mu_{1}}+\frac{\mu_{1}}{2}\right) - e^{\epsilon'}\Phi\left(-\frac{\epsilon'}{\mu_{1}}-\frac{\mu_{1}}{2}\right),
\end{equation}
and $\mu_{1}$ is the solution to the following equation
\begin{equation}
\Phi\left(-\frac{\epsilon}{\mu_{1}}+\frac{\mu_{1}}{2}\right) - e^{\epsilon}\Phi\left(-\frac{\epsilon}{\mu_{1}}-\frac{\mu_{1}}{2}\right) = 0.
\end{equation}
\end{Corollary}
\begin{IEEEproof}
Please see Appendix \ref{CorollaryPrivacy}.
\end{IEEEproof}

\begin{figure}[t]
  \centering
  \centerline{\includegraphics[width=3in]{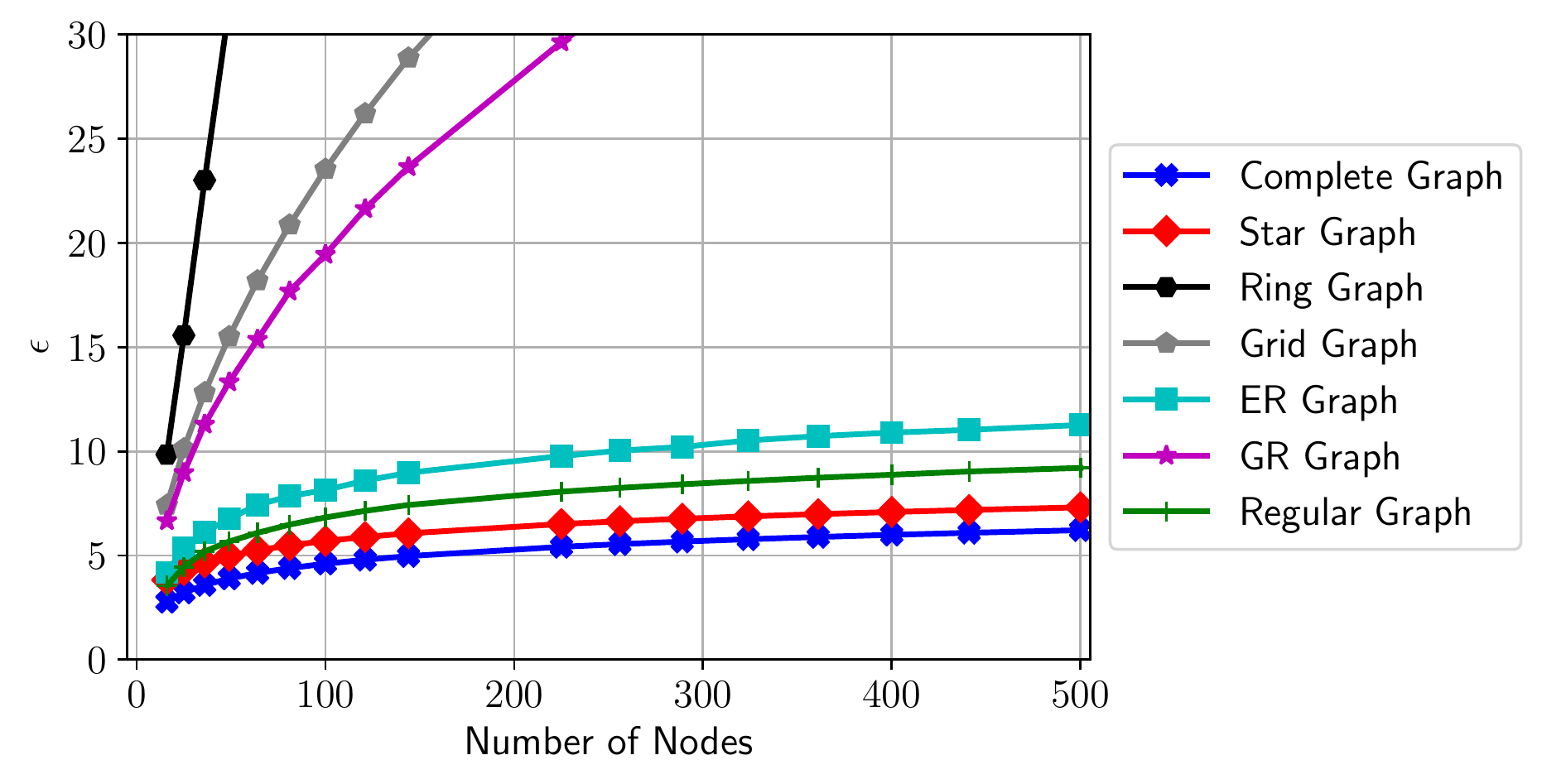}}
\caption{$\epsilon$ v.s. number of nodes ($\alpha = 0.5$)}\label{Epsilon_fig_graphs}
\end{figure}
\begin{figure}[t]
  \centering
  \centerline{\includegraphics[width=3in]{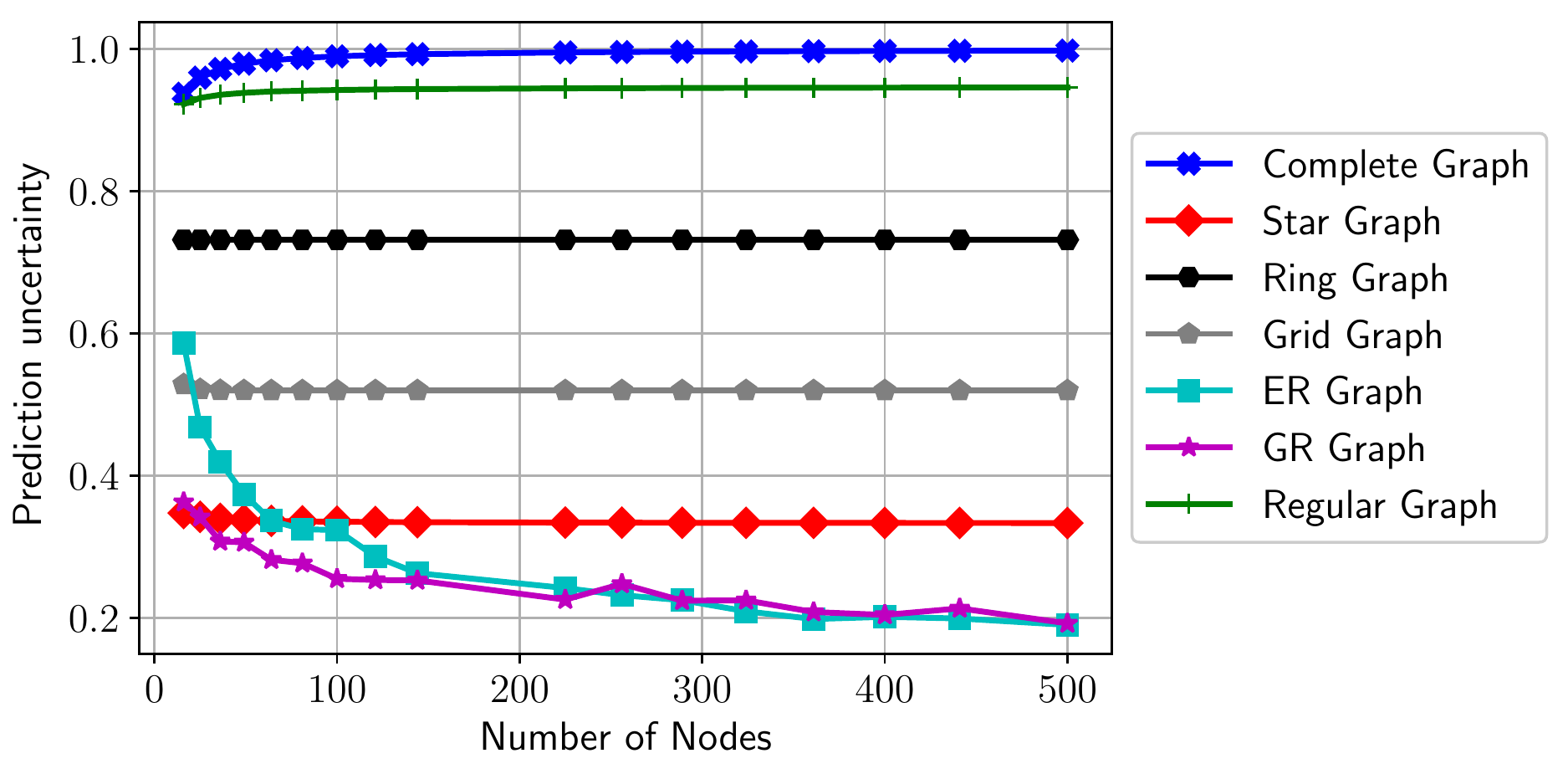}}
\caption{$c$ v.s. number of nodes ($\alpha = 0.5$)}\label{Prediction_fig_graphs}
\end{figure}

Fig. \ref{Epsilon_fig_graphs} and Fig. \ref{Prediction_fig_graphs} show the differential privacy level $\epsilon$ and prediction uncertainty $c$ of Algorithm \ref{PrivateGossip} for several well-known graph topologies, respectively. In particular, in a complete graph, every node is connected to all the other nodes in the graph. A star graph is a graph in which a central node is connected to all the other nodes and the central node is their only neighbor. A ring graph is a graph that consists of a single cycle in which every node has exactly two edges incident with it. For the grid graph, we consider a two-dimensional square grid. These four graphs are deterministic (i.e., given the number of nodes, the graph structures are fixed). We also consider three random graphs, i.e., the regular graph, the Erd\H{o}s R{\'e}nyi (ER) graph and the Geometric Random (GR) graph. An ER graph is a graph in which each node is randomly connected to another node with a certain probability. A GR graph is constructed by randomly placing the nodes in some metric space with the euclidean distance and connecting two nodes by an edge if and only if their distance is less than a specified parameter. A regular graph is a graph in which every node is randomly connected to a fixed number of nodes. In our simulation, we generate these three random graphs using the NetworkX package in Python such that the average degrees are 10.

It can be observed that as the number of node increases, the differential privacy level $\epsilon$ increases for all the graphs. On the one hand, in general, as the number of nodes increases, the probability of each node $i$ being active given the source node $j$ (and therefore $P(j \rightarrow i)$) decreases.\footnote{Note that in the private gossip, only one node is active at each time slot and the active node randomly selects its neighbors to push the message}. On the other hand, the inequality in (\ref{EqM}) should be satisfied for any pair of nodes. That being said, as the number of nodes increases, more inequalities need to be satisfied, which enforces a larger $\epsilon$.

For prediction uncertainty, different graphs exhibit different trends. One possible reason is that for the complete graph and the regular graph, almost all the nodes are the same from the attacker's viewpoint. As a result, as the number of nodes increases, it becomes more difficult to identify the source node. On the other hand, since prediction uncertainty considers the worse case scenario, as the number of nodes increases, the probability that there exists a node different from the other nodes (e.g., with a small degree) increases in the ER graph and the GR graph. As a result, the prediction uncertainty decreases as the number of nodes increases.

\begin{figure}[t]
  \centering
  \centerline{\includegraphics[width=2.4in]{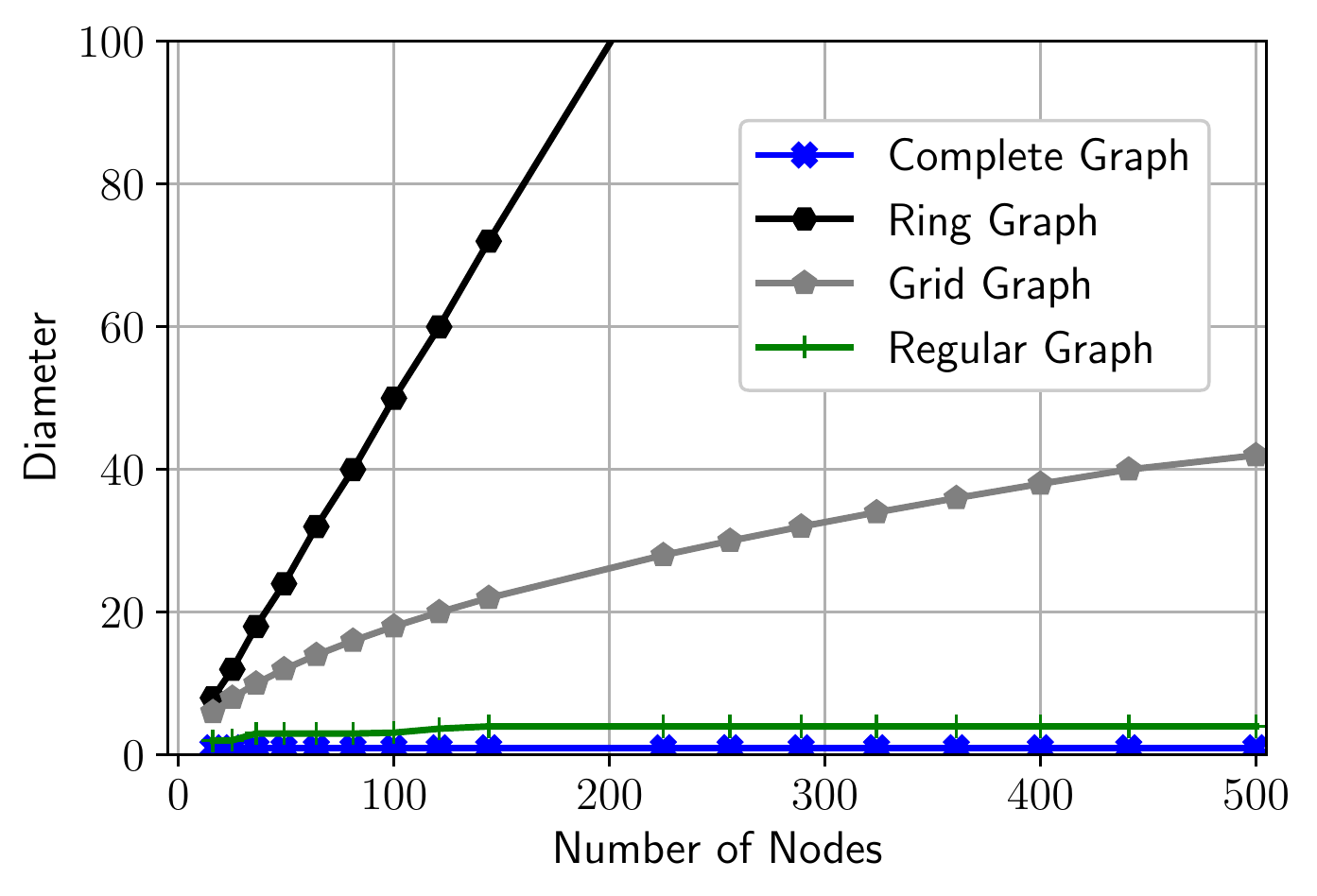}}
\caption{Diameter v.s. number of nodes}\label{diameter_regular_graph}
\end{figure}
\begin{figure}[t]
  \centering
  \centerline{\includegraphics[width=2.4in]{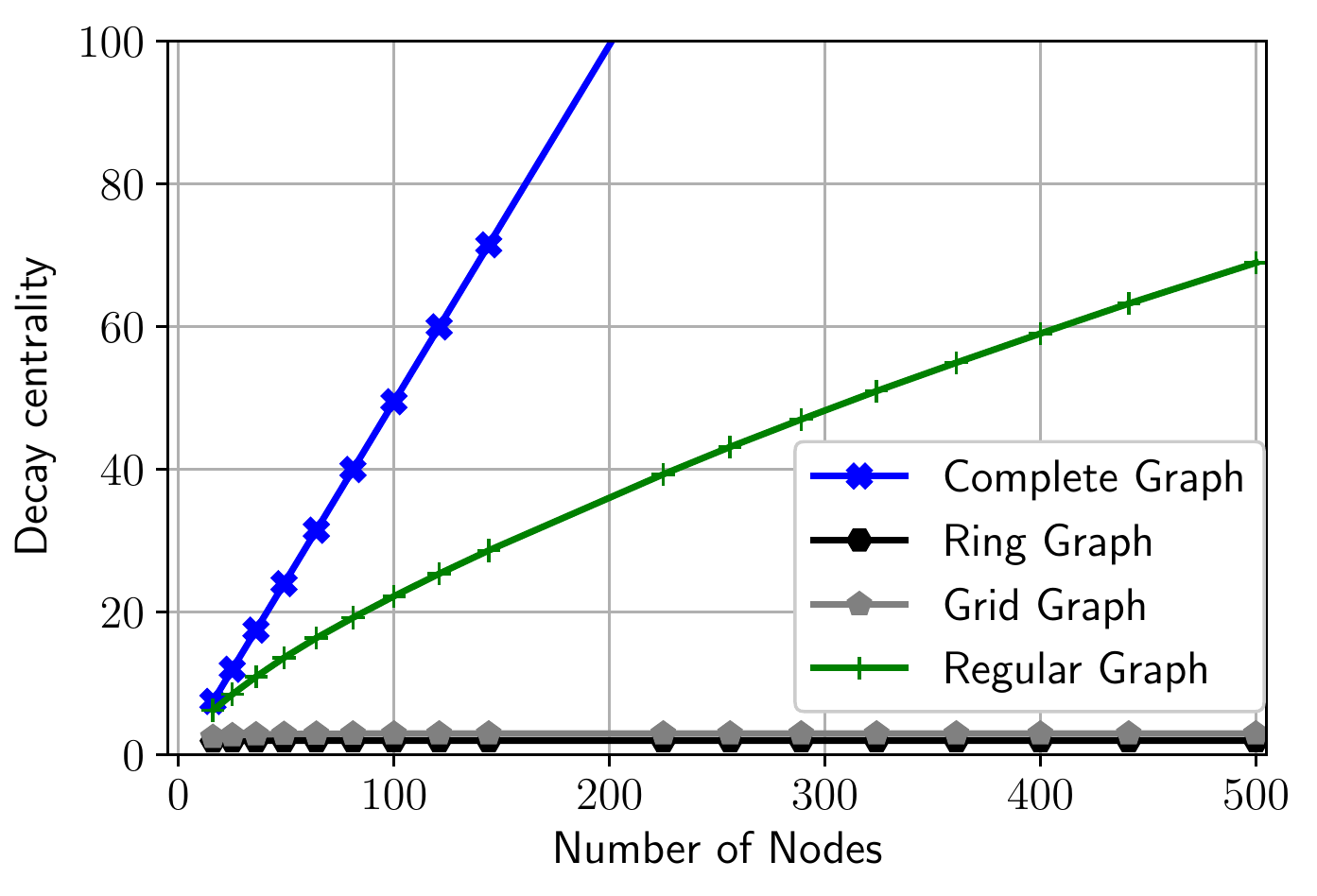}}
\caption{Decay centrality v.s. number of nodes ($\alpha = 0.5$)}\label{decay_regular_graph}
\end{figure}

\begin{figure}[t]
  \centering
  \centerline{\includegraphics[width=2.6in]{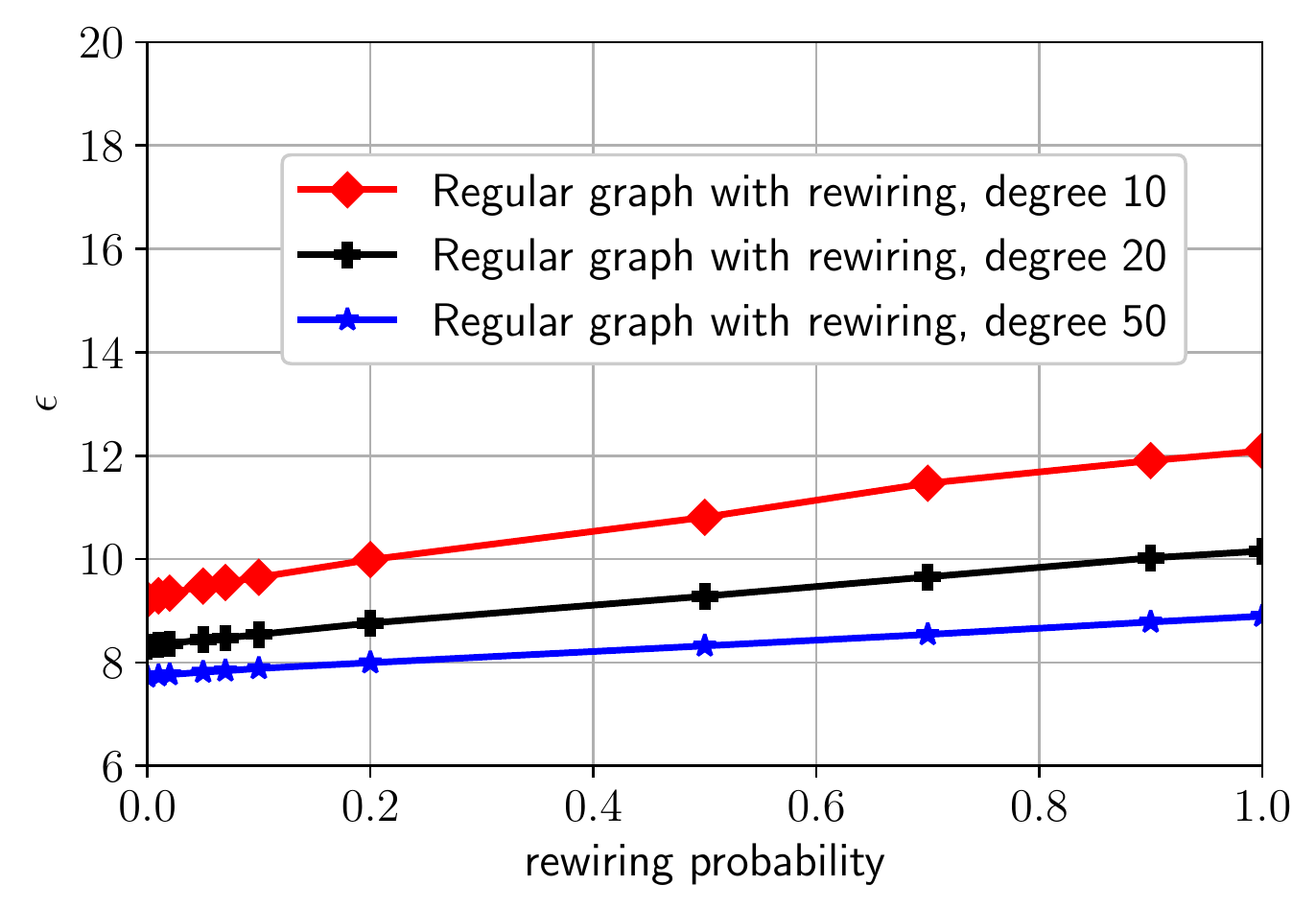}}
\caption{$\epsilon$ v.s. rewiring probability ($n$ = 500, $\alpha = 0.5$)}\label{epsilon_smallworld_fig_graphs}
\end{figure}
\begin{figure}[t]
  \centering
  \centerline{\includegraphics[width=2.6in]{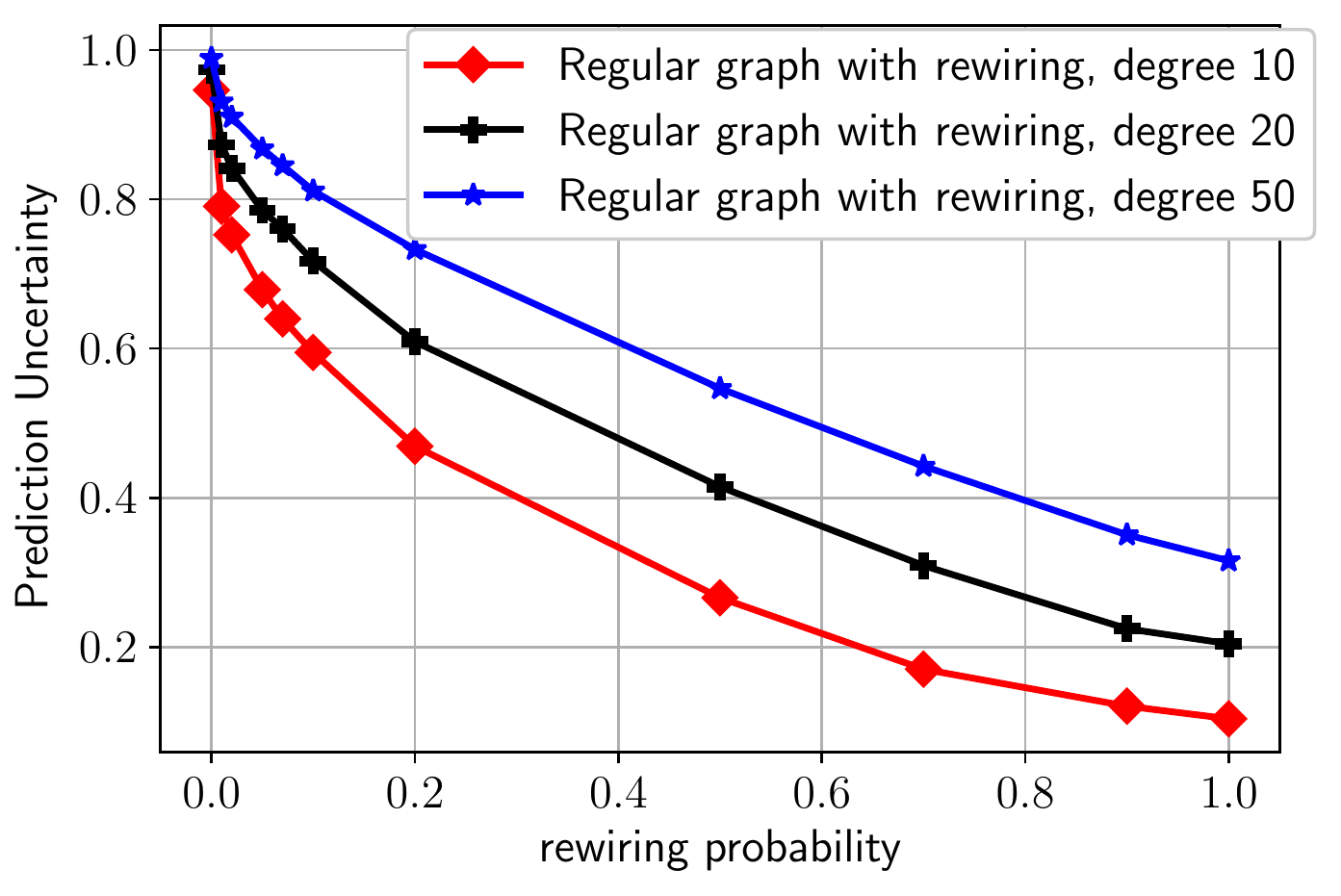}}
\caption{$c$ v.s. rewiring probability ($n$ = 500, $\alpha = 0.5$)}\label{prediction_smallworld_fig_graphs}
\end{figure}

Given the same number of nodes, it can be seen from Fig. \ref{Epsilon_fig_graphs} and Fig. \ref{Prediction_fig_graphs} that the regular graph and the complete graph (a special regular graph with the degree of $n-1$) perform well in terms of both differential privacy and prediction uncertainty. This may be due to the fact that in these graphs every node has the same degree while none of the nodes are distant from the others. As a result, all the nodes look the same from the attacker's perspective. It is worth mentioning that despite that the ring graph and the grid graph can be considered as regular graphs with degrees of 2 and 4, respectively,\footnote{Strictly speaking, the 2-dimensional square grid graph is not regular. However, it is close to a regular graph since most of the nodes have the same degree of 4.} the diameters of these two graphs are large (which also leads to small decay centrality since a large diameter indicates a large shortest path length between the nodes). Therefore, the attacker can still differentiate two distant nodes. Fig. \ref{diameter_regular_graph} and Fig. \ref{decay_regular_graph} show the diameters and the decay centrality of the aforementioned four graphs for a comparison.

To further examine the idea that regular graphs with small diameters have good differential privacy and prediction uncertainty performance, the impact of edge rewiring on the privacy guarantees of regular graphs is examined and the results are presented in Fig. \ref{epsilon_smallworld_fig_graphs} and Fig. \ref{prediction_smallworld_fig_graphs}. Particularly, a random regular graph with 500 nodes and a degree of 10/20/50 is first generated. Then, each edge $(x,y)$ in the graph is replaced by another edge $(x,z)$ with a rewiring probability, where $z$ is randomly sampled from the node set $V$. As the rewiring probability increases, more irregularity is introduced. It can be observed that the differential privacy level $\epsilon$ increases and the prediction uncertainty $c$ decreases, which supports our conjecture.

In summary, given the number of nodes, a graph usually has a smaller $\epsilon$ and a larger $c$ (and therefore better differential privacy and prediction uncertainty) if it has a small diameter and all the nodes have similar degrees.

However, since both differential privacy and prediction uncertainty consider the worst case scenario, they may not necessarily be good measures for source anonymity. To further illustrate this idea, the differential privacy and prediction uncertainty of Algorithm \ref{PrivateGossip} within a candidate set is examined. More specifically, the following theorem can be proved.

\begin{Theorem}\label{candidateepsilon}
Given a general network $G$ and the observation model described in Section \ref{observationmodel} with parameter $\alpha$, Algorithm \ref{PrivateGossip} is $(\epsilon,0)$-differentially private within the candidate set $\mathpzc{Q}$, where
\begin{equation}\label{equationcandidateepsilon}
\begin{split}
\epsilon &= \ln\bigg(\max_{k\notin \mathpzc{Q}, j \neq i \in \mathpzc{Q}}\bigg\{\frac{1}{\alpha(I-(1-\alpha)\hat{A}_{i})^{-1}[j,i]},\\
&~~~~~~~~~~~~~~~~~~~~~~~~~~~~\frac{(I-(1-\alpha)\hat{A}_{k})^{-1}[j,k]}{(I-(1-\alpha)\hat{A}_{k})^{-1}[i,k]}\bigg\}\bigg).
\end{split}
\end{equation}
The prediction uncertainty of Algorithm \ref{PrivateGossip} within the candidate set $\mathpzc{Q}$ is given by
\begin{equation}\label{equationcandidateepsilon2}
\begin{split}
c &= \min_{i\in \mathpzc{Q}, k \notin \mathpzc{Q}}\bigg\{\sum_{j \neq i\in \mathpzc{Q}}\alpha(I-(1-\alpha)\hat{A}_{i})^{-1}[j,i],\\
&~~~~~~~~~~~~~~~~~~~~~~~~\frac{\sum_{j \neq i\in \mathpzc{Q}}\alpha(I-(1-\alpha)\hat{A}_{i})^{-1}[j,k]}{\alpha(I-(1-\alpha)\hat{A}_{i})^{-1}[i,k]}\bigg\}.
\end{split}
\end{equation}
\end{Theorem}
\begin{IEEEproof}
Please see Appendix \ref{Proofcandidateepsilon}.
\end{IEEEproof}

\begin{Remark}
We note that in (\ref{equationcandidateepsilon}) (similarly (\ref{equationcandidateepsilon2})), the first term corresponds to the events for which the attacker observes the activities from the nodes in the candidate set before those from the nodes outside the candidate set. Therefore, it is similar to that in Theorem \ref{dpprivategossip}. The second term corresponds to the events for which attacker first observes the activities from the nodes outside the candidate set.
\end{Remark}

\begin{figure}[t]
  \centering
  \centerline{\includegraphics[width=2.6in]{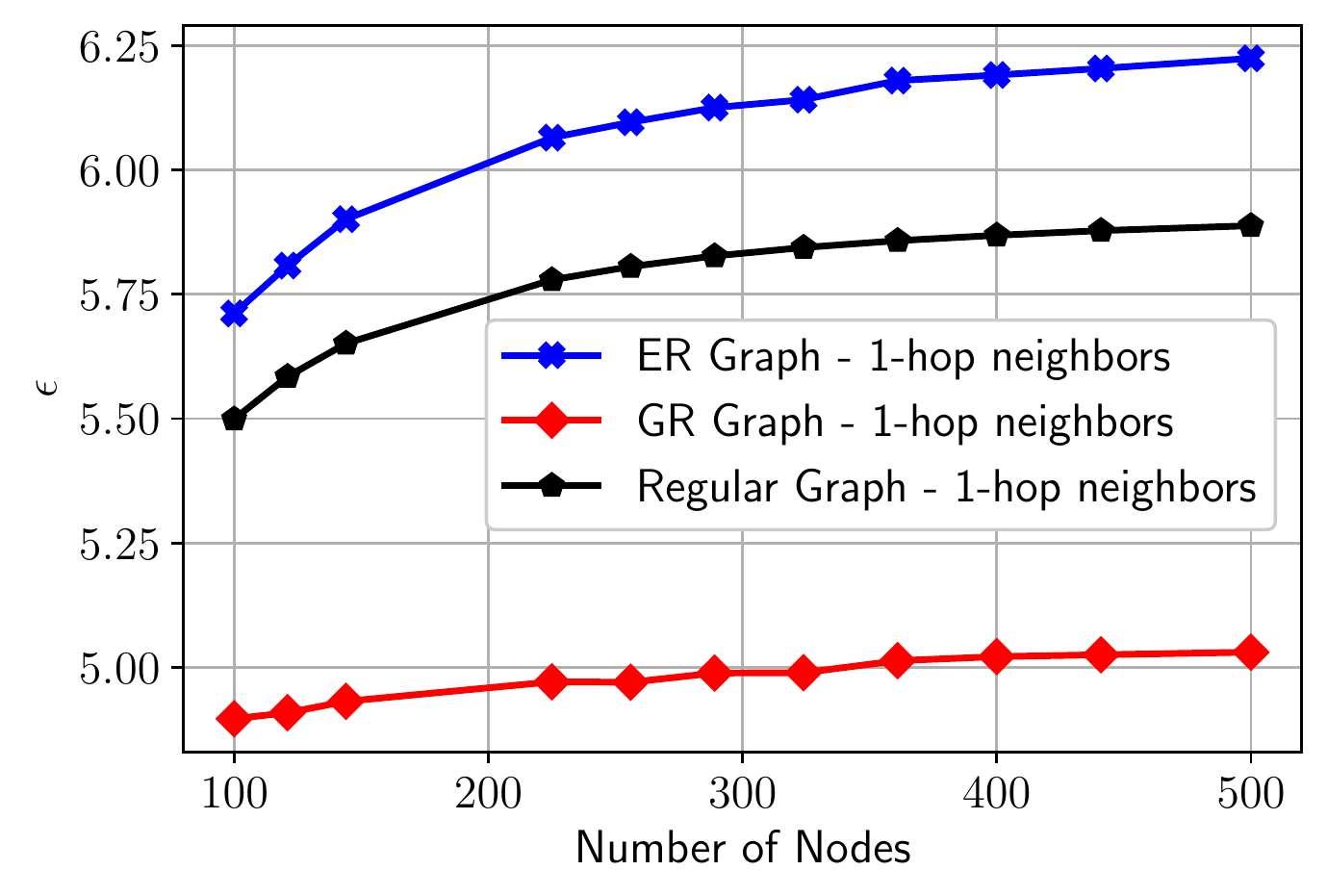}}
\caption{$\epsilon$ v.s. number of nodes ($\alpha = 0.5$)}\label{epsilon_fig_candidate_graphs}
\end{figure}
\begin{figure}[t]
  \centering
  \centerline{\includegraphics[width=2.6in]{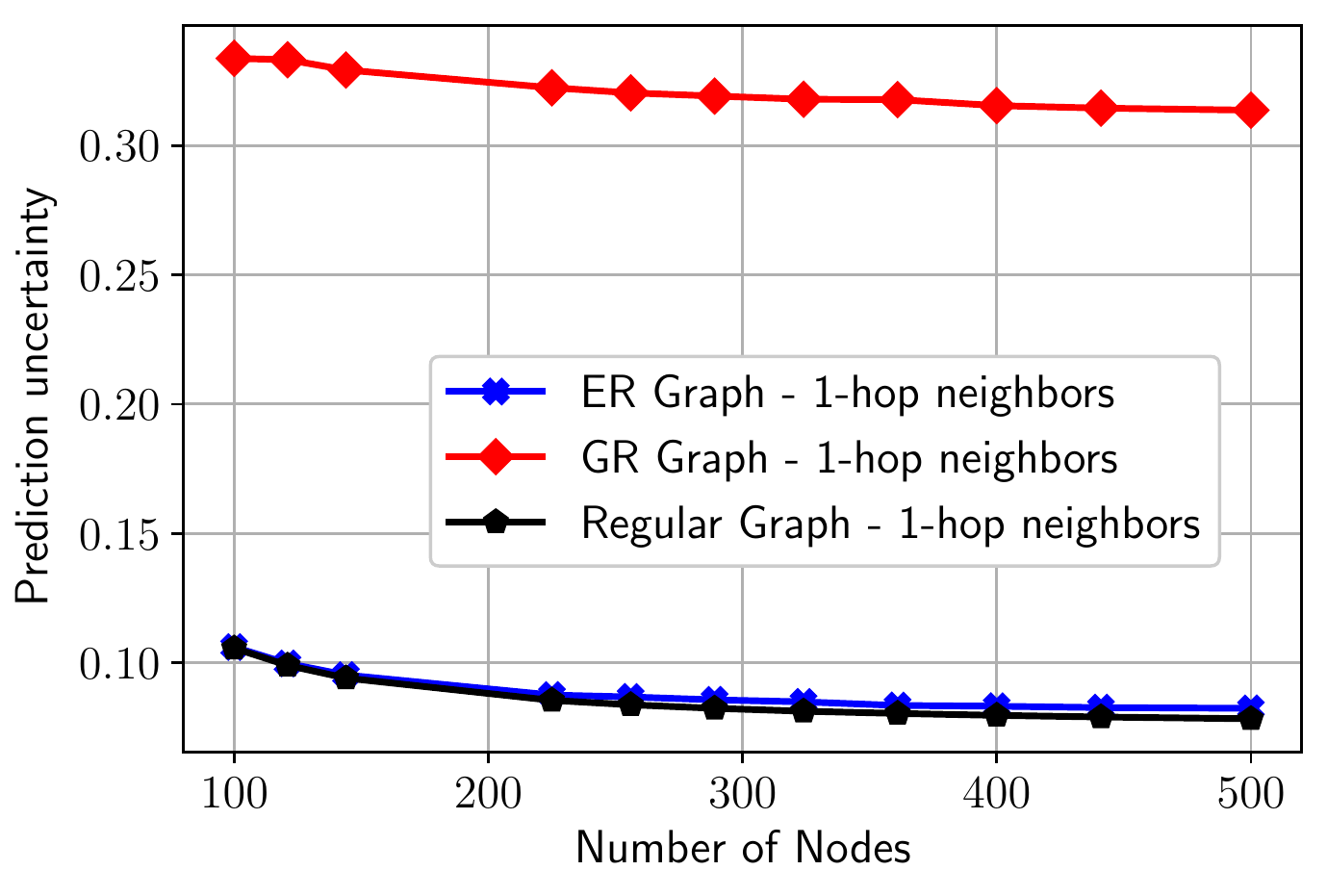}}
\caption{$c$ v.s. number of nodes ($\alpha = 0.5$)}\label{prediction_fig_candidate_graphs}
\end{figure}

Fig. \ref{epsilon_fig_candidate_graphs} and Fig. \ref{prediction_fig_candidate_graphs} show the differential privacy level $\epsilon$ and prediction uncertainty $c$ of Algorithm \ref{PrivateGossip} for the three random graphs within a candidate set, respectively. The average degrees of all the graphs are set as 10. In particular, the candidate set $\mathpzc{Q}$ contains a randomly selected source node and its 1-hop neighbors. 15,000 Monte Carlo runs are performed for each graph, and the average $\epsilon$ and $c$ within the candidate set $\mathpzc{Q}$ are presented.

Different from the results in Fig. \ref{Epsilon_fig_graphs} and Fig. \ref{Prediction_fig_graphs}, the GR graph performs better than the ER graph and the regular graph. This is because, in the GR graph, there exist some clusters that are well connected locally. In addition, the 1-hop neighboring nodes of a node is likely to be in the same cluster. As a result, it is more difficult for the attacker to distinguish two nodes in the candidate set for the GR graph.

\begin{Remark}
The above results show that the original differential privacy and prediction uncertainty may have limitations in measuring the privacy guarantees of gossip protocols. Particularly, Fig. \ref{Epsilon_fig_graphs} and Fig. \ref{Prediction_fig_graphs} show that the regular graph has a smaller differential privacy level $\epsilon$ and a larger prediction uncertainty $c$ than the GR graph. However, it does not necessarily mean that the regular graph always provides better privacy protection. Fig. \ref{epsilon_fig_candidate_graphs} and Fig. \ref{prediction_fig_candidate_graphs} show that the GR graph performs better than the regular graph in terms of differential privacy and prediction uncertainty within the 1-hop neighbors. In practice, the attacker can often obtain some side information about the source. For instance, the source of the leaked information about a company usually has close connection to the company. In this case, the source may not be interested to hide itself among all the nodes in the whole network, but instead among those closely related to the company. That being said, if the source node cares more about hiding itself among a subset of nodes (e.g., its 1-hop neighbors), differential privacy and prediction uncertainty within the corresponding candidate set may serve as better privacy metrics.

\end{Remark}

\begin{figure}[t]
  \centering
  \centerline{\includegraphics[width=2.6in]{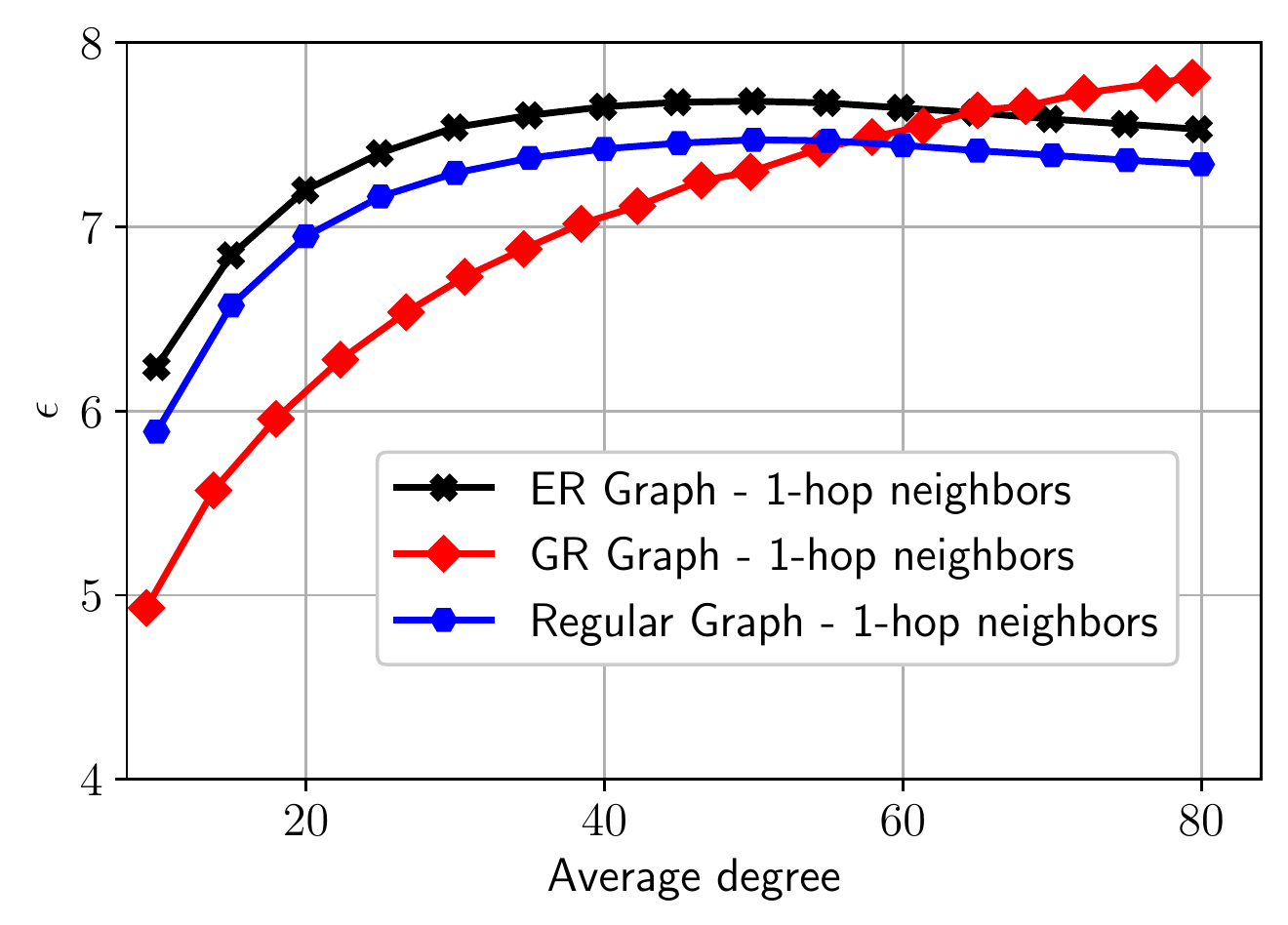}}
\caption{$\epsilon$ v.s. average degree ($n=500$, $\alpha = 0.5$)}\label{epsilon_fig_candidate_degree_graphs}
\end{figure}
\begin{figure}[t]
  \centering
  \centerline{\includegraphics[width=2.6in]{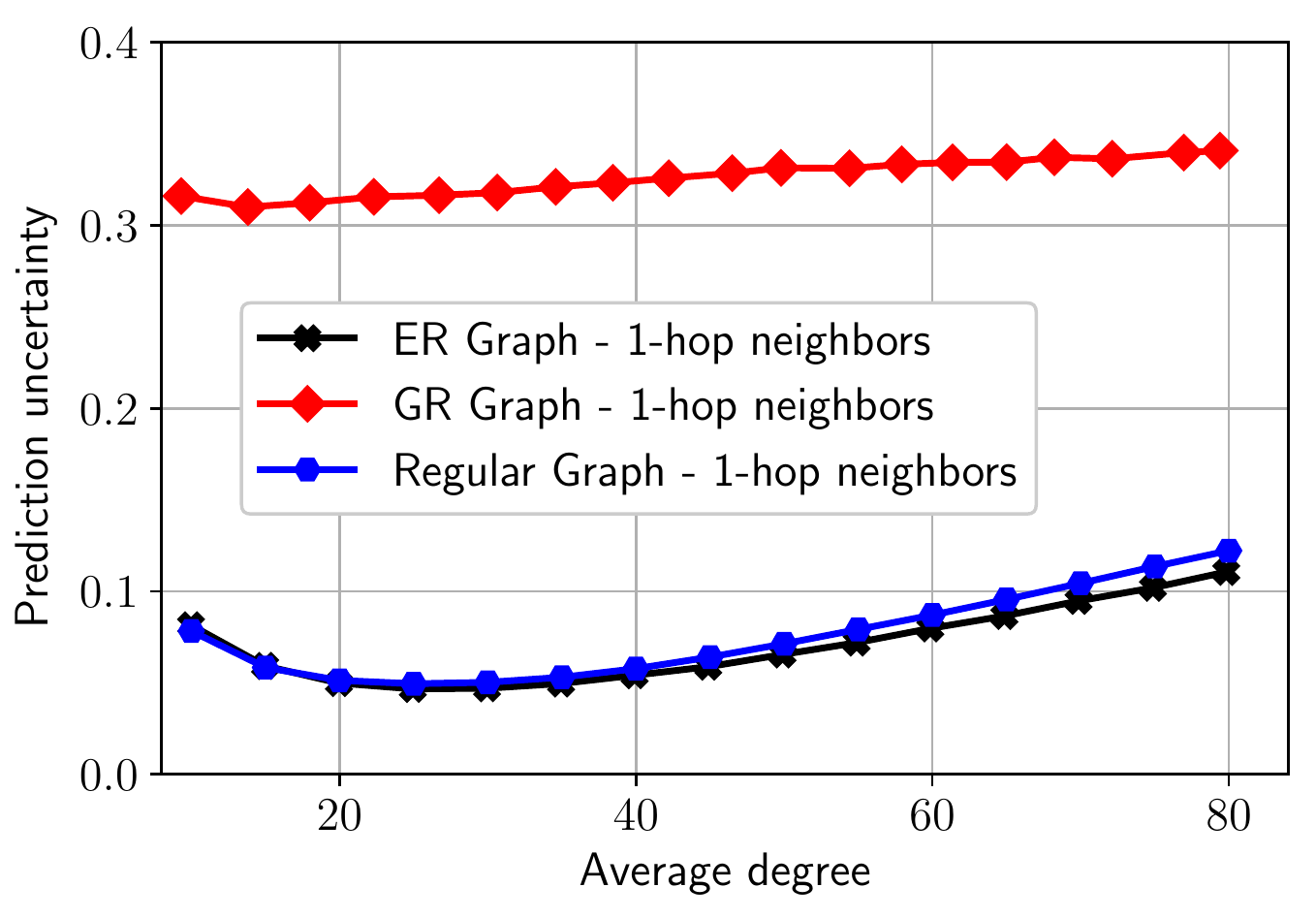}}
\caption{$c$ v.s. average degree ($n=500$, $\alpha = 0.5$)}\label{prediction_fig_candidate_degree_graphs}
\end{figure}

Fig. \ref{epsilon_fig_candidate_degree_graphs} and Fig. \ref{prediction_fig_candidate_degree_graphs} show the impact of average degree on $\epsilon$ and $c$ for the three random graphs. It can be observed that as the average degree increases, $\epsilon$'s of the ER graph and the regular graph first increase and then decrease. Intuitively, $\epsilon$ depends on two parameters: the size of the candidate set and the connectivity of the graphs. As the average degree increases, the size of the candidate set becomes larger, while the graphs become better connected. When the average degree is small, the impact of the candidate set size dominates that of the connectivity. As a result, increasing the average degree leads to a larger $\epsilon$. When the average degree is large enough, the impact of the connectivity dominates that of the candidate set size. Therefore, a larger average degree corresponds to a smaller $\epsilon$. The same analysis can be applied to prediction uncertainty. On the other hand, for the GR graph, as the average degree increases, the number of nodes in the candidate set increases. Since the nodes in the candidate set are usually well connected (like a complete graph), it is similar to the complete graph with an increasing number of nodes. As a result, the trends of the curves for the GR graph are similar to those of the complete graph in Fig. \ref{Epsilon_fig_graphs} and Fig. \ref{Prediction_fig_graphs}.

\section{Privacy-Spreading Tradeoff of Gossip Protocols in Wireless Networks}\label{tradeoffprivacyspreading}
\noindent Considering that in many real world applications, the information spreading between two nodes may be realized through wireless communications \cite{zhang2013gossip,dimakis2008geographic}, the privacy guarantees of gossip protocols in wireless networks are investigated in this subsection. It is assumed that the communications between the network nodes and between the attacker and its deployed sensors are prone to errors due to various interferences. To simplify the analysis, a failure probability is considered in this setting: Due to interferences, the communications will fail with a probability of $f$ between two nodes during the $gossip$ step, and it is assumed that the attacker fails to receive a report from any of its deployed sensors about the detected events with the same probability $f$.\footnote{ As a first work in this area, this simplified assumption is adopted to facilitate the characterization of the tradeoff between privacy and spreading speed. More realistic assumption concerning two different but correlated failure probabilities \cite{vuran2006spatial} warrants further study.} Note that the failure probability $f$, induced by detrimental effects in wireless channels, is different from the detection probability $\alpha$ that is due to the limitation in the eavesdropping capability (e.g., computation power) of the sensors. In this case, the privacy guarantees of gossip protocols are characterized in the following theorem.

\begin{Theorem}\label{TW}
Considering the same setting as in Theorem \ref{T1}, with the additional constraint that both the legitimate communication and the adversarial reporting fail with a probability $f$, the gossip-based protocols can guarantee $(\epsilon, \delta)$-differential privacy with
$\delta\geq \alpha(1-f)$ and $c$-prediction uncertainty with $c=0$
in the synchronous setting, and
 $\delta\geq max[\alpha(1-f)-e^{\epsilon}(1-\alpha(1-f))^{D_G}, \alpha(1-f)-e^{\epsilon}\frac{1-\alpha(1-f)}{n-1}]$ and $c\leq \min\limits_{i\in V}\frac{C_{1-\alpha(1-f)}(i)}{\alpha(1-f)}$ in the asynchronous setting.
\end{Theorem}

The privacy guarantees of Algorithm \ref{PrivateGossip} is characterized in the following theorem.

\begin{Theorem}
Consider the same setting as in Theorem \ref{dpprivategossip}, with the additional constraint that both the legitimate communication and the adversarial reporting fail with a probability $f$, Algorithm \ref{PrivateGossip} is $(\epsilon,0)$-differentially private in the asynchronous setting, where
\begin{equation}
\epsilon = \ln\bigg(\max_{j \neq i \in V}\frac{1}{\alpha(1-f)(I-(1-\alpha(1-f))\hat{A}_{i})^{-1}[j,i]}\bigg).
\end{equation}
The prediction uncertainty of Algorithm \ref{PrivateGossip} in the asynchronous setting is given by
\begin{equation}
c = \min_{i\in V}\sum_{j \neq i\in V}\alpha(1-f)(I-(1-\alpha(1-f)\hat{A}_{i}))^{-1}[j,i].
\end{equation}
\end{Theorem}

The proofs of the above theorems readily follow from the previous results and the details are omitted in the interest of space.

Adding artificial noise is a typical way to enhance privacy in practical applications \cite{abadi2016deep}. In wireless networks, interference is a natural source for privacy enhancement as it hampers the attacker's observations of the network activities, which can be further strengthened through approaches such as friendly jamming \cite{vilela2011wireless}. However, the information spreading process is impeded as well in such scenarios. The information spreading time of the standard and the private gossip protocols in this case is given below.
\begin{Theorem}\label{Theorem6}
In a wireless network $G$ in which the communications fail with a probability of $f$, we have
\begin{enumerate}
\item In the synchronous setting, the private gossip takes $C_G/(1-f)$ rounds on average to inform all nodes in the network, where $C_G$ is the cover time of a random walk in network $G$.
\item In the asynchronous setting, the private gossip takes $C_G/(1-f)$ time on average, while the standard gossip takes $T_{as}/(1-f)$ time on average to finish spreading, where $T_{as}$ is the spreading time of standard gossip when the communication is perfect.
\end{enumerate}
\end{Theorem}
\begin{IEEEproof}
Please see Appendix \ref{Prooftheorem6}.
\end{IEEEproof}

\begin{figure}[t]
  \centering
  \centerline{\includegraphics[width=2.6in]{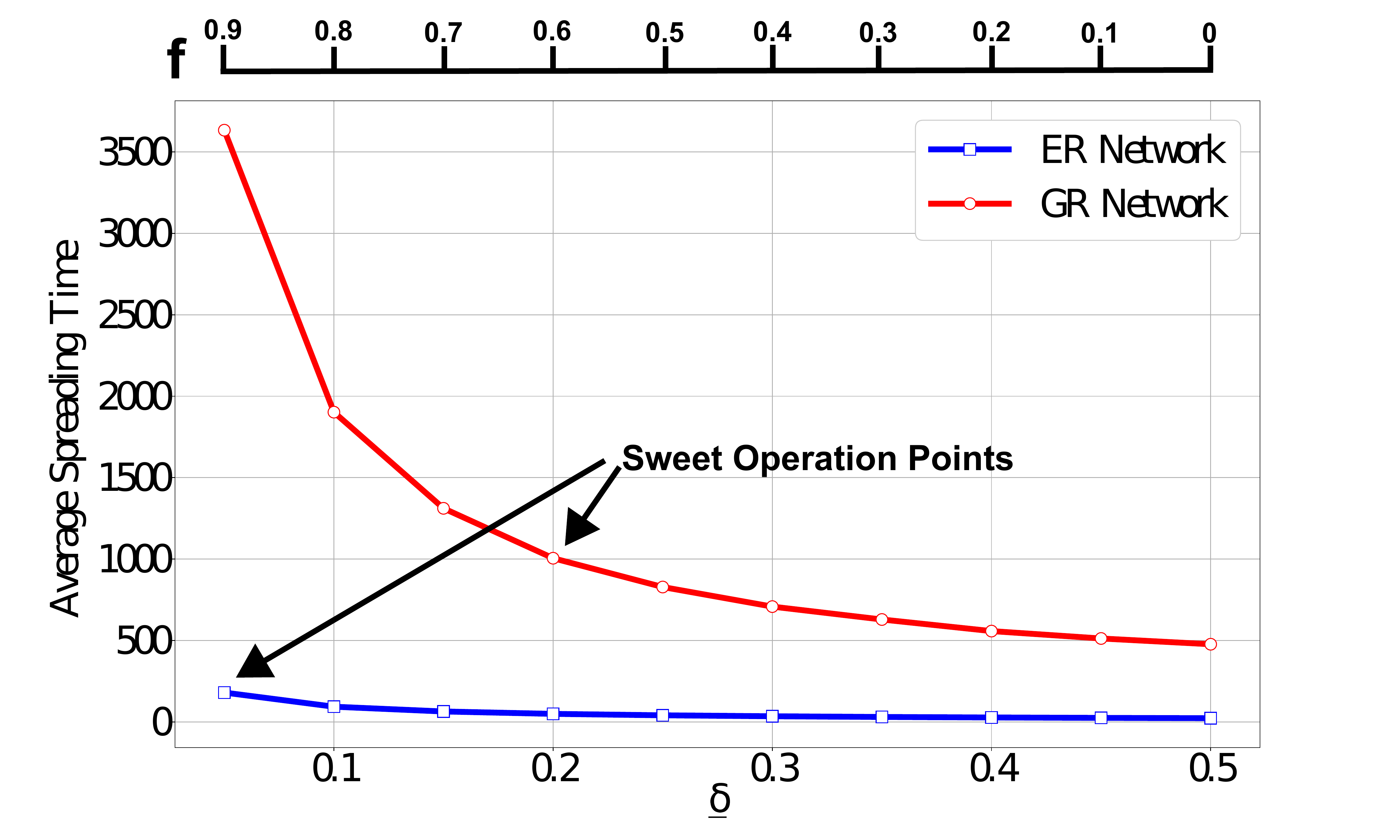}}
\caption{DP v.s. spreading speed in the synchronous setting}\label{F1}
\vspace{-0.2in}
\end{figure}

\begin{figure}[t]
  \centering
  \centerline{\includegraphics[width=2.6in]{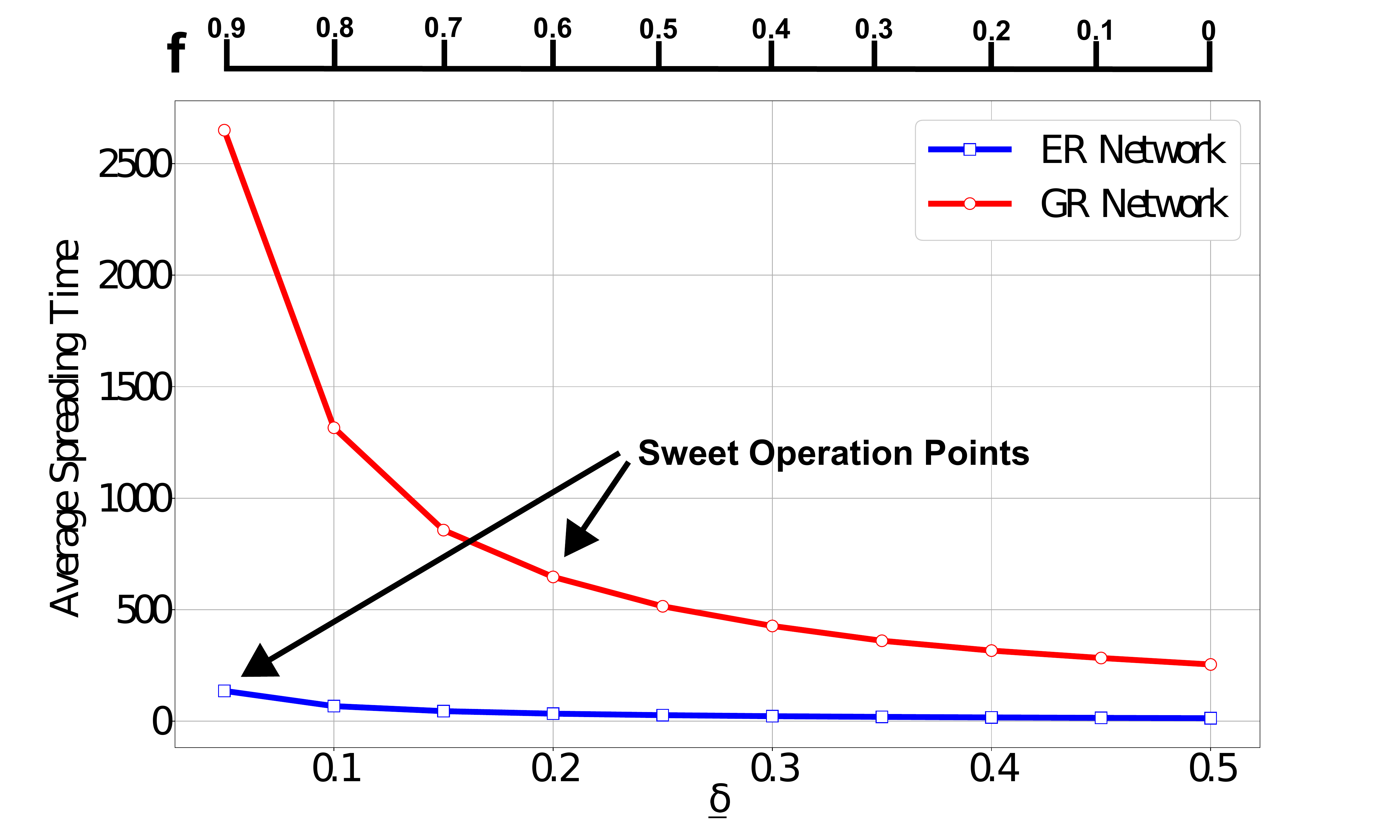}}
\caption{DP v.s. spreading speed in the asynchronous setting}\label{F2}
\vspace{-0.2in}
\end{figure}

\begin{Remark}
For standard gossip in the synchronous setting, multiple random walks can exist during the spreading process, which renders the analysis of unreliable spreading challenging in general networks. But we conjecture that a similar result as in the synchronous setting may hold.

The above results indicate a trade-off between privacy and spreading speed of gossip protocols, which is further explored through simulations below. In particular, following the existing literature in information spreading (e.g., \cite{min2016layer,picu2012analysis}), ER networks and GR Networks with a total number of $n=100000$ nodes and average node degree of $10$ are considered. Each point in the following figures is obtained through simulations with $5$ network instances and $100$ Monte Carlo runs for each instance. The average $90\%$ spreading time is considered \cite{zhang2013gossip}. The privacy-spreading tradeoffs for ER and GR networks for standard gossip in the synchronous and asynchronous settings are shown in Fig. \ref{F1} and Fig. \ref{F2}, respectively. It is assumed that $\alpha = 0.5$ and privacy budget $\epsilon = 1$ without loss of generality. The corresponding privacy lower bounds $\underline{\delta}$ in the x-axis are calculated for the considered ER and GR networks using Theorem \ref{TW} given the failure probability $f$ (one-to-one correspondence). Similar results are obtained for private gossip and omitted here due to the space constraint.
\end{Remark}


\begin{Remark}
Through analysis, it can be seen that the spreading time is inversely proportional to $1-f$ while the privacy lower bound $\underline{\delta}$ is proportional to $1-f$. From Fig. \ref{F1} and Fig. \ref{F2}, it can be seen that when $\underline{\delta}$ increases from 0.05 to 0.1, for GR networks, the average spreading time decreases from around 3600 and 2600 to 1800 and 1300 in the synchronous and asynchronous settings, respectively. This means that we can trade a small loss of privacy for dramatic improvement in spreading time. On the other hand, for ER networks or GR networks with large $\underline{\delta}$ (small $f$), the average spreading time increases slowly as $\underline{\delta}$ decreases. Therefore, the privacy guarantees of gossip protocols can be strengthened with a small loss of spreading time (e.g., the sweet operation points in Fig. \ref{F1} and Fig. \ref{F2}), which suggests that methods like adding artificial noise can be useful in privacy-preserving information spreading.
\end{Remark}

\section{Privacy of Gossip Protocols in Delayed Monitoring}\label{delatedminitoringprivacy}
\noindent In reality, the attacker may not monitor the whole information spreading process right from the beginning. In this section, we try to quantify the differential privacy of general gossip protocols when the monitoring is delayed. To avoid complication, it is assumed that the communications between nodes and the reception at the attacker are perfect. In addition, the attacker knows the global time in the synchronous setting or the number of communication that has occurred in the asynchronous setting since the beginning of information spreading.

\begin{Theorem}\label{delayedgeneral}
Considering the same setting as in Theorem \ref{T1}, if the attacker starts monitoring the information spreading process $t$ rounds (or $t$ steps of gossip communications in the asynchronous case) after it begins and $t < D_G$, the gossip-based protocols can guarantee $(\epsilon, \delta)$-differential privacy with $\delta\geq \frac{1}{d_{max}^t}\alpha$ in the synchronous setting. In the asynchronous setting
\begin{equation}\label{LEq}
\begin{aligned}
\delta \geq &\max\bigg[\frac{1}{d_{max}^t (t+1)!}\alpha-e^{\epsilon}(1-\alpha)^{D_G-t}, \\ &\frac{1}{d_{max}^t (t+1)!}\alpha-e^{\epsilon}\frac{1-\frac{1}{d_{max}^t (t+1)!}\alpha}{n-1}, 0\bigg],
\end{aligned}
\end{equation}
in which $d_{max}=max_{i\in V}d_{i}$ is the largest node degree.
\end{Theorem}
\begin{IEEEproof}
Please see Appendix \ref{Proofdelayedgeneral}.
\end{IEEEproof}
\begin{Remark}
Gossip protocols are not able to protect the source's identity effectively during the early stage of information spreading. As the spreading process continues, more and more randomness is introduced, leading to stronger and stronger privacy. Therefore, in delayed monitoring, it becomes more difficult for the attacker to identify the source node as the delay increases.
\end{Remark}

\begin{Theorem}\label{delayedepsilon}
Given a general network $G$ and the observation model described in Section \ref{observationmodel} with parameter $\alpha$, if the attacker starts monitoring the information spreading process $t$ steps of gossip communications after it begins, Algorithm \ref{PrivateGossip} is $(\epsilon,0)$-differentially private in the asynchronous setting, where
\begin{equation}\label{delayedepsilonequation}
\small
\begin{split}
&e^{\epsilon}=\\
&\max_{i\in V, j \neq z \in V}\frac{A^{t}[j,i]+\sum_{k\neq i \in V}A^{t}[j,k]\alpha (I-(1-\alpha)\hat{A}_{i})^{-1}[k,i]}{A^{t}[z,i]+\sum_{k\neq i \in V}A^{t}[z,k]\alpha (I-(1-\alpha)\hat{A}_{i})^{-1}[k,i]}.
\end{split}
\end{equation}
The prediction uncertainty of Algorithm \ref{PrivateGossip} in the asynchronous setting is given by
\begin{equation}
\small
\begin{split}
&c = \min_{i \in V, z \in V}\bigg\{\\
&\frac{\sum_{j\neq i \in V}[A^{t}[j,z]+\sum_{k\neq z \in V}A^{t}[j,k]\alpha (I-(1-\alpha)\hat{A}_{z})^{-1}[k,z]]}{A^{t}[i,z]+\sum_{k\neq z \in V}A^{t}[i,k]\alpha (I-(1-\alpha)\hat{A}_{z})^{-1}[k,z]}\bigg\}
\end{split}
\end{equation}
\end{Theorem}
\begin{IEEEproof}
Please see Appendix \ref{Proofdelayedepsilon}.
\end{IEEEproof}

\begin{Remark}
Note that in (\ref{delayedepsilonequation}), $\alpha (I-(1-\alpha)\hat{A}_{i})^{-1}[k,i] = P(k \rightarrow i) = \alpha \sum_{m=0}^{\infty}(1-\alpha)^{m}\hat{A}_{i}^{m}[k,i]$. For any $k \neq i, \hat{A}_{i}^{0}[k,i] = I[k,i] = 0$. Therefore,
\begin{equation}
\begin{split}
\alpha (I-(1-\alpha)\hat{A}_{i})^{-1}[k,i] &= \alpha \sum_{m=1}^{\infty}(1-\alpha)^{m}\hat{A}_{i}^{m}[k,i] \\
&\leq \alpha \sum_{m=1}^{\infty}(1-\alpha)^{m} = 1- \alpha.
\end{split}
\end{equation}
As a result,
\begin{equation}
\begin{split}
&A^{t}[j,i]+\sum_{k\neq i \in V}A^{t}[j,k]\alpha (I-(1-\alpha)\hat{A}_{i})^{-1}[k,i] \\
&< A^{t}[j,i]+\sum_{k\neq i \in V}A^{t}[j,k] = 1.
\end{split}
\end{equation}
On the other hand,
\begin{equation}
\begin{split}
&A^{t}[z,i]+\sum_{k\neq i \in V}A^{t}[z,k]\alpha (I-(1-\alpha)\hat{A}_{i})^{-1}[k,i] \\
&> A^{t}[z,i]\min_{k\neq i \in V}\alpha(I-(1-\alpha)\hat{A}_{i})^{-1}[k,i]\\
&+ \sum_{k\neq i \in V}A^{t}[z,k]\alpha(I-(1-\alpha)\hat{A}_{i})^{-1}[k,i]\\
&\geq \min_{k\neq i \in V}\alpha(I-(1-\alpha)\hat{A}_{i})^{-1}[k,i].
\end{split}
\end{equation}
Comparing Eq. (\ref{delayedepsilonequation}) with Eq. (\ref{epsilonstandard}), we can see that the $\epsilon$ in Theorem \ref{delayedepsilon} is smaller than that in Theorem \ref{dpprivategossip}, i.e., in delayed monitoring, the differential privacy guarantee is enhanced. Similar result can be obtained for prediction uncertainty.
\end{Remark}

\begin{figure}[t]
  \centering
  \centerline{\includegraphics[width=2.6in]{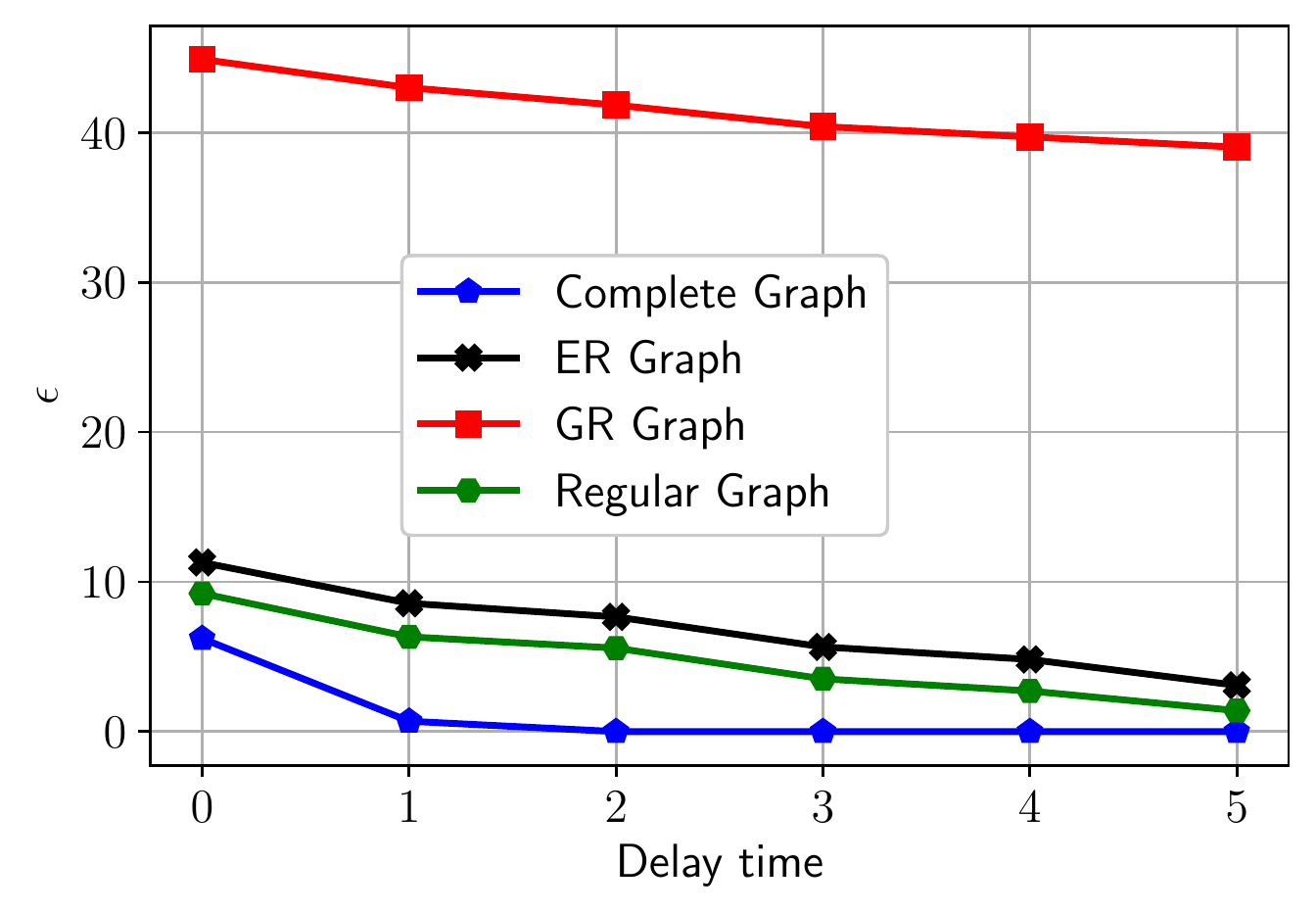}}
\caption{$\epsilon$ v.s. delay time ($n = 500$, $\alpha = 0.5$)}\label{epsilon_fig_delay_graphs}
\end{figure}
\begin{figure}[t]
  \centering
  \centerline{\includegraphics[width=2.6in]{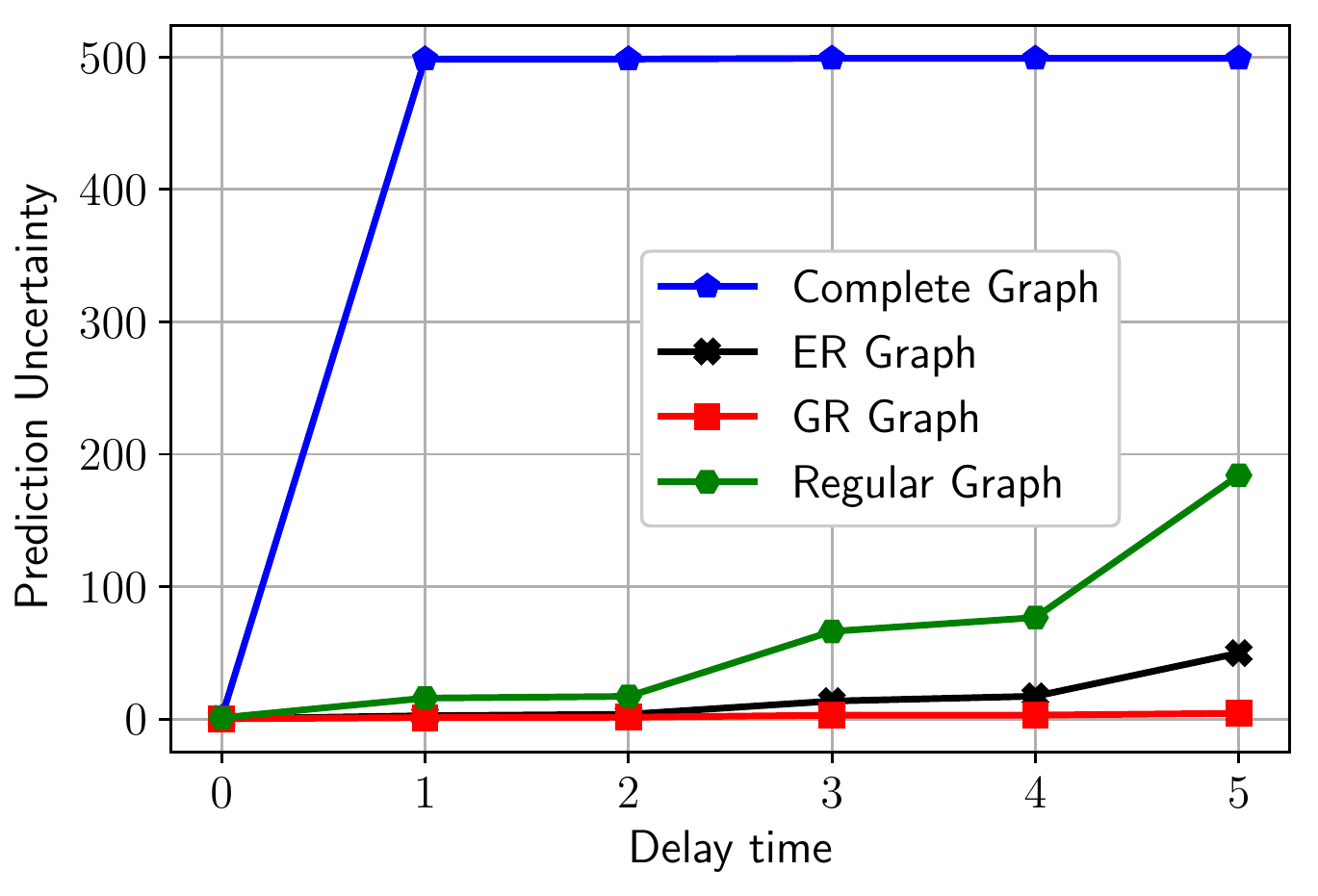}}
\caption{$c$ v.s. delay time ($n = 500$, $\alpha = 0.5$)}\label{prediction_fig_delay_graphs}
\end{figure}

Fig. \ref{epsilon_fig_delay_graphs} and Fig. \ref{prediction_fig_delay_graphs} show the impact of delay time $t$ on the differential privacy level $\epsilon$ and the prediction uncertainty $c$, respectively. It can be observed that as the delay time increases, $\epsilon$ ($c$) quickly decreases (increases) for the ER graph, the regular graph and the complete graph. This is because, in these graphs, within the delay time $t$, the information is quickly spread to the rest of the nodes in the graph, which makes it difficult for the attacker to identify the source. For the GR graph, however, during the delay time, it is likely that the information remains in the same cluster that the source node belongs to. As a result, it is still not difficult for the attacker to distinguish the nodes across clusters.

\section{Conclusions and Future Works}\label{CF}
\noindent In this work, the privacy guarantees of gossip-based protocols in general networks are investigated. Bounds on differential privacy and prediction uncertainty of gossip protocols in general networks are obtained. For the private gossip algorithm, the differential privacy and prediction uncertainty guarantees are derived in closed form. It is found that these two metrics are closely related to some key network structure parameters, such as network diameter and decay centrality, in the (arguably) more interesting asynchronous setting. Moreover, considering that differential privacy and prediction uncertainty may be restrictive in some scenarios, the relaxed variants of these two metrics are proposed. In wireless networks, through a simplified modeling for unreliable communications, the tradeoff between privacy and spreading efficiency is revealed, and it is suggested that natural or artificial interference can enhance the privacy of gossip protocols with the cost of a decrease in spreading speed. Finally, in delayed monitoring, it is verified that the privacy of gossip protocols is enhanced as the delayed time increases, and the corresponding effect is quantified.

Many interesting problems remain open in this line of research besides those already mentioned above. Investigating more appropriate privacy metrics and designing more effective privacy-aware gossip protocols for different observation models (e.g., network snapshot \cite{jiang2016identifying}) and network structures are interesting future directions.



\bibliographystyle{IEEEtran} 
\bibliography{refs}

\begin{thebibliography}{10}
\providecommand{\url}[1]{#1}
\csname url@samestyle\endcsname
\providecommand{\newblock}{\relax}
\providecommand{\bibinfo}[2]{#2}
\providecommand{\BIBentrySTDinterwordspacing}{\spaceskip=0pt\relax}
\providecommand{\BIBentryALTinterwordstretchfactor}{4}
\providecommand{\BIBentryALTinterwordspacing}{\spaceskip=\fontdimen2\font plus
\BIBentryALTinterwordstretchfactor\fontdimen3\font minus
  \fontdimen4\font\relax}
\providecommand{\BIBforeignlanguage}[2]{{%
\expandafter\ifx\csname l@#1\endcsname\relax
\typeout{** WARNING: IEEEtran.bst: No hyphenation pattern has been}%
\typeout{** loaded for the language `#1'. Using the pattern for}%
\typeout{** the default language instead.}%
\else
\language=\csname l@#1\endcsname
\fi
#2}}
\providecommand{\BIBdecl}{\relax}
\BIBdecl

\bibitem{chandra2001anonymous}
R.~Chandra, V.~Ramasubramanian, and K.~Birman, ``Anonymous gossip: Improving
  multicast reliability in mobile ad-hoc networks,'' in \emph{Proceedings 21st
  International Conference on DCS}, 2001, pp. 275--283.

\bibitem{dimakis2008geographic}
A.~D. Dimakis, A.~D. Sarwate, and M.~J. Wainwright, ``Geographic gossip:
  Efficient averaging for sensor networks,'' \emph{IEEE Transactions on Signal
  Processing}, vol.~56, no.~3, pp. 1205--1216, 2008.

\bibitem{ganesh2003peer}
A.~J. Ganesh, A.-M. Kermarrec, and L.~Massouli{\'e}, ``Peer-to-peer membership
  management for gossip-based protocols,'' \emph{IEEE transactions on
  computers}, vol.~52, no.~2, pp. 139--149, 2003.

\bibitem{bianchi2013convergence}
P.~Bianchi and J.~Jakubowicz, ``Convergence of a multi-agent projected
  stochastic gradient algorithm for non-convex optimization,'' \emph{IEEE
  Transactions on Automatic Control}, vol.~58, no.~2, pp. 391--405, 2013.

\bibitem{liu2018differentially}
Y.~Liu, J.~Liu, and T.~Basar, ``Differentially private gossip gradient
  descent,'' in \emph{Proceedings of the 57th IEEE CDC}, 2018, pp. 2777--2782.

\bibitem{jiang2016identifying}
J.~Jiang, S.~Wen, S.~Yu, Y.~Xiang, and W.~Zhou, ``Identifying propagation
  sources in networks: State-of-the-art and comparative studies,'' \emph{IEEE
  Communications Surveys \& Tutorials}, vol.~19, no.~1, pp. 465--481, 2016.

\bibitem{fanti2017hiding}
G.~Fanti, P.~Kairouz, S.~Oh, K.~Ramchandran, and P.~Viswanath, ``Hiding the
  rumor source,'' \emph{IEEE Transactions on Information Theory}, vol.~63,
  no.~10, pp. 6679--6713, 2017.

\bibitem{dwork2014algorithmic}
C.~Dwork, A.~Roth \emph{et~al.}, ``The algorithmic foundations of differential
  privacy,'' \emph{Foundations and Trends{\textregistered} in Theoretical
  Computer Science}, vol.~9, no. 3--4, pp. 211--407, 2014.

\bibitem{bellet2019started}
A.~Bellet, R.~Guerraoui, and H.~Hendrikx, ``Who started this rumor? quantifying
  the natural differential privacy guarantees of gossip protocols,''
  \emph{arXiv preprint arXiv:1902.07138}, 2019.

\bibitem{backstrom2012four}
L.~Backstrom, P.~Boldi, M.~Rosa, J.~Ugander, and S.~Vigna, ``Four degrees of
  separation,'' in \emph{Proceedings of the 4th Annual ACM Web Science
  Conference}, 2012, pp. 33--42.

\bibitem{jackson2010social}
M.~O. Jackson, \emph{Social and economic networks}.\hskip 1em plus 0.5em minus
  0.4em\relax Princeton university press, 2010.

\bibitem{zhang2013gossip}
H.~Zhang, Z.~Zhang, and H.~Dai, ``Gossip-based information spreading in mobile
  networks,'' \emph{IEEE Transactions on Wireless Communications}, vol.~12,
  no.~11, pp. 5918--5928, 2013.

\bibitem{vuran2006spatial}
M.~C. Vuran and I.~F. Akyildiz, ``Spatial correlation-based collaborative
  medium access control in wireless sensor networks,'' \emph{IEEE/ACM
  Transactions On Networking}, vol.~14, no.~2, pp. 316--329, 2006.

\bibitem{abadi2016deep}
M.~Abadi, A.~Chu, I.~Goodfellow, H.~B. McMahan, I.~Mironov, K.~Talwar, and
  L.~Zhang, ``Deep learning with differential privacy,'' in \emph{Proceedings
  of the 2016 ACM SIGSAC Conference on Computer and Communications Security},
  2016, pp. 308--318.

\bibitem{vilela2011wireless}
J.~P. Vilela, M.~Bloch, J.~Barros, and S.~W. McLaughlin, ``Wireless secrecy
  regions with friendly jamming,'' \emph{IEEE Transactions on Information
  Forensics and Security}, vol.~6, no.~2, pp. 256--266, 2011.

\bibitem{min2016layer}
B.~Min, S.-H. Gwak, N.~Lee, and K.-I. Goh, ``Layer-switching cost and
  optimality in information spreading on multiplex networks,'' \emph{Scientific
  reports}, vol.~6, p. 21392, 2016.

\bibitem{picu2012analysis}
A.~Picu, T.~Spyropoulos, and T.~Hossmann, ``An analysis of the information
  spreading delay in heterogeneous mobility dtns,'' in \emph{2012 IEEE
  International Symposium on WoWMoM}, 2012, pp. 1--10.

\bibitem{dong2019gaussian}
J.~Dong, A.~Roth, and W.~J. Su, ``Gaussian differential privacy,'' \emph{arXiv
  preprint arXiv:1905.02383}, 2019.

\end{thebibliography}

\newpage
\begin{center}
{\Large \textbf{Appendices}}
\end{center}
\appendices
\section{Proof of Theorem \ref{T1}}\label{ProofTheoremT1}
\begin{IEEEproof}
First, for the synchronous setting, let $\mathpzc{S}_{i,0}$ be the event that node $i$'s activity is observed by the attacker's sensors at time $0$. Then, the probability that such an event happens given the source node is $i$ is $p_G^{(i)}(\mathpzc{S}_{i,0})=\alpha$. If the source node is any other node $j\neq i$, $p_G^{(j)}(\mathpzc{S}_{i,0})=0$ since node $i$ cannot initialize a communication if it is not a source node at time $0$. Therefore, $\delta\geq \alpha$ and $c=0$.

In the asynchronous setting, let $\mathpzc{S}_{i,0}$ be the event that node $i$'s activity is observed by the attacker's sensors as its first observed event. It can be seen that, if the source node is $i$, then $p_G^{(i)}(\mathpzc{S}_{i,0})= p_G^{(i)}(\mathpzc{S}_{i,0}|T_{i,0})p_G^{(i)}(T_{i,0})+p_G^{(i)}(\mathpzc{S}_{i,0}|\overline{T}_{i,0})p_G^{(i)}(\overline{T}_{i,0})\geq\alpha$, where $T_{i,0}$ stands for the event that the source node is detected during its first communication. If the source node is $j$, we can consider the following event, denoted as $O_{d(i,j)}$: there is no communication detected by the sensors in the network after $d(i,j)$ gossip actions have been executed from the beginning. Then we have
\begin{equation}\label{EqFP}
\begin{aligned}
p_G^{(j)}(\mathpzc{S}_{i,0})&=p_G^{(j)}(\mathpzc{S}_{i,0}\bigcap\overline{O}_{d(i,j)})+p_G^{(j)}(\mathpzc{S}_{i,0}\bigcap O_{d(i,j)})\\
&=p_G^{(j)}(\mathpzc{S}_{i,0}\bigcap O_{d(i,j)})\\
&\leq p_G^{(j)}(O_{d(i,j)})=(1-\alpha)^{d(i,j)},
\end{aligned}
\end{equation}
where the second equality is due to the fact that $\mathpzc{S}_{i,0}\bigcap\overline{O}_{d(i,j)} = \emptyset$, as it takes at least $d(i,j)$ communications for the information to be delivered to node $i$ from node $j$. Since $p_G^{(i)}(\mathpzc{S}_{i,0})\geq \alpha$, by applying Lemma 1, we have
\begin{equation}\label{EqD}
\begin{aligned}
\delta &\geq\max_{i,j}(\alpha - e^{\epsilon}(1-\alpha)^{d(i,j)})= \alpha - e^{\epsilon}(1-\alpha)^{D_G}.
\end{aligned}
\end{equation}
On the other hand, since $\sum_{j\in V}p_G^{(i)}(\mathpzc{S}_{j,0})=1$, there exists a node $l\in V$ such that
\begin{equation}\label{EqP}
\begin{aligned}
p_G^{(i)}(\mathpzc{S}_{l,0}) &\leq \frac{1}{n-1}\sum_{j\in V, j\neq i}p_G^{(i)}(\mathpzc{S}_{j,0})\\
&= \frac{1 - p_G^{(i)}(\mathpzc{S}_{i,0})}{n-1}\leq \frac{1-\alpha}{n-1}.\\
\end{aligned}
\end{equation}

This implies $\delta \geq \alpha - e^{\epsilon}\frac{1-\alpha}{n-1}$. By Eq. (\ref{EqD}), we have
\begin{equation}\label{Tolerence1}
\delta \geq \max\left[\alpha-e^{\epsilon}(1-\alpha)^{D_G}, \alpha-e^{\epsilon}\frac{1-\alpha}{n-1}\right].
\end{equation}
Meanwhile, as we have $p_G^{(j)}(\mathpzc{S}_{i,0})\leq (1-\alpha)^{d(i,j)}$, the detection uncertainty can be calculated as
\begin{equation}
\begin{aligned}
c= \min_{i,\mathpzc{S}}\left(\frac{\sum_{j\neq i}p_G^{(j)}(\mathpzc{S})}{p_G^{(i)}(\mathpzc{S})}\right)&\leq \min_{i}\frac{\sum_{j\neq i}(1-\alpha)^{d(i,j)}}{\alpha}\\
&=\min_{i\in V}\frac{C_{1-\alpha}(i)}{\alpha}.
\end{aligned}
\end{equation}
\end{IEEEproof}

\section{Proof of Lemma \ref{TheoremProbability}}\label{ProofTheoremProbability}
\begin{IEEEproof}
Let $P_{m}(j \rightarrow i)$ denote the probability that node $i$ is active at time $m$ for the first time before the attacker observes the first event given the source node $j$. Then we have
\begin{equation}\label{probability}
P(j \rightarrow i) = \sum_{m=1}^{\infty}P_{m}(j \rightarrow i).
\end{equation}
Recall from Remark \ref{remark5} that the probability of node $i$ being active at time $m$ for the first time is $\hat{A}_{i}^{m}[j,i] - \hat{A}_{i}^{m-1}[j,i]$. Therefore, we have
\begin{equation}\label{mstepprobability}
P_{m}(j \rightarrow i) = (1-\alpha)^{m}[\hat{A}_{i}^{m}[j,i] - \hat{A}_{i}^{m-1}[j,i]].
\end{equation}
Plugging (\ref{mstepprobability}) into (\ref{probability}) yields
\begin{equation}\label{equation32}
\begin{split}
P(j \rightarrow i) &= \sum_{m=1}^{\infty}(1-\alpha)^{m}[\hat{A}_{i}^{m}[j,i] - \hat{A}_{i}^{m-1}[j,i]] \\
&=\sum_{m=1}^{\infty}(1-\alpha)^{m}\hat{A}_{i}^{m}[j,i] - \sum_{m=1}^{\infty}(1-\alpha)^{m}\hat{A}_{i}^{m-1}[j,i]\\
&= \sum_{m=1}^{\infty}(1-\alpha)^{m}\hat{A}_{i}^{m}[j,i] - \sum_{m=1}^{\infty}(1-\alpha)^{m+1}\hat{A}_{i}^{m}[j,i] \\
&- (1-\alpha)\hat{A}_{i}^{0}[j,i] \\
&=\sum_{m=1}^{\infty}\alpha (1-\alpha)^{m}\hat{A}_{i}^{m}[j,i] - (1-\alpha)I[j,i] \\
&= \alpha \sum_{m=0}^{\infty}(1-\alpha)^{m}\hat{A}_{i}^{m}[j,i].
\end{split}
\end{equation}
Since $\hat{A}_{i}$ is a right stochastic matrix, it is known that its eigenvalues $|\lambda_{\hat{A}_{i}}|\leq 1$. Therefore, the eigenvalues of $(1-\alpha)\hat{A}_{i}$ are strictly less than 1. As a result, (\ref{equation32}) can be written alternatively as
\begin{equation}
\begin{split}
P(j \rightarrow i) &= \alpha \sum_{m=0}^{\infty}(1-\alpha)^{m}\hat{A}_{i}^{m}[j,i] \\
&= \alpha(I-(1-\alpha)\hat{A}_{i})^{-1}[j,i],
\end{split}
\end{equation}
which completes the proof.
\end{IEEEproof}

\section{Proof of Theorem \ref{dpprivategossip}}\label{Proofdpprivategossip}
\begin{IEEEproof}
Let $\mathpzc{S}_{i,0}$ denote the event such that node $i$'s activity is observed by the attacker as its first observation. Then, for any $j \neq i$ we have
\begin{equation}
p^{j}_{G}(\mathpzc{S}_{i,0}) = P(j \rightarrow i)p_{G}^{i}(\mathpzc{S}_{i,0}) \leq p_{G}^{i}(\mathpzc{S}_{i,0}).
\end{equation}
On the other hand, by Lemma \ref{TheoremProbability}
\begin{equation}
\begin{split}
p^{i}_{G}(\mathpzc{S}_{i,0}) &= \frac{1}{P(j \rightarrow i)}p_{G}^{j}(\mathpzc{S}_{i,0}) \\
&= \frac{1}{\alpha(I-(1-\alpha)\hat{A}_{i})^{-1}[j,i]}p_{G}^{j}(\mathpzc{S}_{i,0}).
\end{split}
\end{equation}
Therefore, according to Definition \ref{definition1}
\begin{equation}\label{eepsilon}
e^{\epsilon} = \max_{j \neq i \in V}\frac{1}{\alpha(I-(1-\alpha)\hat{A}_{i})^{-1}[j,i]}.
\end{equation}
Taking logarithm on both sides of (\ref{eepsilon}) gives the results in Theorem \ref{dpprivategossip}.

Regarding the prediction uncertainty, we have
\begin{equation}
\begin{split}
&\frac{p_G(I_0\neq\{i\}|\mathpzc{S}_{i,0})}{p_G(I_0=\{i\}|\mathpzc{S}_{i,0})}\\
& =\frac{\sum_{j \neq i \in V}p_G^{(j)}(\mathpzc{S}_{i,0})}{p_G^{(i)}(\mathpzc{S}_{i,0})}\\
&=\frac{\sum_{j \neq i \in V}\alpha(I-(1-\alpha)\hat{A}_{i})^{-1}[j,i] p_G^{(i)}(\mathpzc{S}_{i,0})}{p_G^{(i)}(\mathpzc{S}_{i,0})}\\
&=\sum_{j \neq i \in V}\alpha(I-(1-\alpha)\hat{A}_{i})^{-1}[j,i].
\end{split}
\end{equation}
For any $k \neq i \in V$, it can be shown that
\begin{equation}
\begin{split}
&\frac{p_G(I_0\neq\{i\}|\mathpzc{S}_{k,0})}{p_G(I_0=\{i\}|\mathpzc{S}_{k,0})}\\
&=\frac{\sum_{j \neq i \in V}p_G^{(j)}(\mathpzc{S}_{k,0})}{p_G^{(i)}(\mathpzc{S}_{k,0})}\\
&=\frac{p_{G}^{(k)}(\mathpzc{S}_{k,0})+\sum_{j \neq i,k \in V}p_G^{(j)}(\mathpzc{S}_{k,0})}{p_G^{(i)}(\mathpzc{S}_{k,0})}\\
&=\frac{p_{G}^{(k)}(\mathpzc{S}_{k,0})+\sum_{j \neq i,k \in V}\alpha(I-(1-\alpha)\hat{A}_{k})^{-1}[j,k]p_G^{(k)}(\mathpzc{S}_{k,0})}{\alpha(I-(1-\alpha)\hat{A}_{k})^{-1}[i,k]p_G^{(k)}(\mathpzc{S}_{k,0})}\\
&=\frac{1+\sum_{j \neq i,k \in V}\alpha(I-(1-\alpha)\hat{A}_{k})^{-1}[j,k]}{\alpha(I-(1-\alpha)\hat{A}_{k})^{-1}[i,k]}\\
&\geq \frac{\sum_{j \neq k \in V}\alpha(I-(1-\alpha)\hat{A}_{k})^{-1}[j,k]}{\alpha(I-(1-\alpha)\hat{A}_{k})^{-1}[i,k]}\\
&\geq \sum_{j \neq k \in V}\alpha(I-(1-\alpha)\hat{A}_{k})^{-1}[j,k]\\
&=\frac{p_G(I_0\neq\{k\}|\mathpzc{S}_{k,0})}{p_G(I_0=\{k\}|\mathpzc{S}_{k,0})}.
\end{split}
\end{equation}
As a result.
\begin{equation}
\begin{split}
c &= \min_{i,\mathpzc{S}\subseteq \mathbb{S}}\left(\frac{p_G(I_0\neq\{i\}|\mathpzc{S})}{p_G(I_0=\{i\}|\mathpzc{S})}\right) \\
&= \min_{i}\sum_{j \neq i \in V}\alpha(I-(1-\alpha)\hat{A}_{i})^{-1}[j,i],
\end{split}
\end{equation}
which completes the proof.
\end{IEEEproof}

\section{Proof of Corollary \ref{corollarydp}}\label{CorollaryPrivacy}
\begin{IEEEproof}
The proof of Corollary \ref{corollarydp} readily follows from the following results on Gaussian differential privacy.
\begin{Theorem}\cite{dong2019gaussian}
A mechanism is $\mu$-Gaussian differentially private if and only if it is $(\epsilon, \delta(\epsilon))$-differentially private for all $\epsilon \geq 0$, where
\begin{equation}\label{equation40}
\delta(\epsilon) = \Phi\left(-\frac{\epsilon}{\mu}+\frac{\mu}{2}\right) - e^{\epsilon}\Phi\left(-\frac{\epsilon}{\mu}-\frac{\mu}{2}\right).
\end{equation}
\end{Theorem}
Given any $(\epsilon,0)$-differentially private mechanism, the corresponding $\mu_{1}$ can be obtained by (\ref{equation40}). Applying $\mu_{1}$ and the above theorem completes the proof. Interested readers may refer to \cite{dong2019gaussian} for more details.
\end{IEEEproof}

\section{Proof of Theorem \ref{candidateepsilon}}\label{Proofcandidateepsilon}
\begin{IEEEproof}
Let $\mathpzc{S}_{i,0}$ denote the event such that node $i$'s activity is observed by the attacker as its first observation. For any $j \neq i \in \mathpzc{Q}$, similar to the proof of Theorem \ref{dpprivategossip}, we have
\begin{equation}
p^{(j)}_{G}(\mathpzc{S}_{i,0}) = \alpha (I-(1-\alpha)\hat{A}_{i})^{-1}[j,i]p^{(i)}_{G}(\mathpzc{S}_{i,0}) \leq p^{(i)}_{G}(\mathpzc{S}_{i,0}),
\end{equation}
and
\begin{equation}
p^{(i)}_{G}(\mathpzc{S}_{i,0}) = \frac{1}{\alpha (I-(1-\alpha)\hat{A}_{i})^{-1}[j,i]}p^{(j)}_{G}(\mathpzc{S}_{i,0}).
\end{equation}
For any $k \notin \mathpzc{Q}$ and $j \neq i \in \mathpzc{Q}$, we have
\begin{equation}
p^{(j)}_{G}(\mathpzc{S}_{k,0}) = \alpha (I-(1-\alpha)\hat{A}_{k})^{-1}[j,k]p^{(k)}_{G}(\mathpzc{S}_{k,0}),
\end{equation}
and
\begin{equation}
p^{(i)}_{G}(\mathpzc{S}_{k,0}) = \alpha (I-(1-\alpha)\hat{A}_{k})^{-1}[i,k]p^{(k)}_{G}(\mathpzc{S}_{k,0}),
\end{equation}
Therefore,
\begin{equation}
p^{(j)}_{G}(\mathpzc{S}_{k,0}) = \frac{\alpha (I-(1-\alpha)\hat{A}_{k})^{-1}[j,k]}{\alpha (I-(1-\alpha)\hat{A}_{k})^{-1}[i,k]}p^{(i)}_{G}(\mathpzc{S}_{k,0}).
\end{equation}
As a result,
\begin{equation}\label{eepsiloncandidate}
\begin{split}
e^{\epsilon} &= \max_{k\notin \mathpzc{Q}, j \neq i \in \mathpzc{Q}}\bigg\{\frac{1}{\alpha(I-(1-\alpha)\hat{A}_{i})^{-1}[j,i]},\\
&~~~~~~~~~~~~~~~~~~~~~~~~~~~~\frac{(I-(1-\alpha)\hat{A}_{k})^{-1}[j,k]}{(I-(1-\alpha)\hat{A}_{k})^{-1}[i,k]}\bigg\}.
\end{split}
\end{equation}
Taking logarithm on both sides of (\ref{eepsiloncandidate}) gives the result in Theorem \ref{candidateepsilon}.

Regarding the prediction uncertainty, recall that for any $\mathpzc{S} \subseteq \mathbb{S}$, we have
\begin{equation}
\begin{split}
&\frac{p_G(I_0\neq\{i\}|\mathpzc{S})}{p_G(I_0=\{i\}|\mathpzc{S})}=\frac{\sum_{j \neq i \in \mathpzc{Q}}p_G^{(j)}(\mathpzc{S})}{p_G^{(i)}(\mathpzc{S})}.
\end{split}
\end{equation}
Similar to the proof of Theorem \ref{dpprivategossip}, for any event $\mathpzc{S}_{i,0} \subseteq \mathbb{S}$ and $j\neq i \in \mathpzc{Q}$, it can be shown that
\begin{equation}
\begin{split}
&\frac{p_G(I_0\neq\{i\}|\mathpzc{S}_{i,0})}{p_G(I_0=\{i\}|\mathpzc{S}_{i,0})}\\
&=\frac{\sum_{j \neq i \in \mathpzc{Q}}p_G^{(j)}(\mathpzc{S}_{i,0})}{p_G^{(i)}(\mathpzc{S}_{i,0})}\\
&=\frac{\sum_{j \neq i \in \mathpzc{Q}}\alpha(I-(1-\alpha)\hat{A}_{i})^{-1}[j,i]p_G^{(i)}(\mathpzc{S}_{i,0})}{p_G^{(i)}(\mathpzc{S}_{i,0})}\\
&=\sum_{j \neq i \in \mathpzc{Q}}\alpha(I-(1-\alpha)\hat{A}_{i})^{-1}[j,i].
\end{split}
\end{equation}
For any event $\mathpzc{S}_{k,0} \subseteq \mathbb{S}$ such that $k \in \mathpzc{Q}$, 
\begin{equation}
\frac{p_G(I_0\neq\{i\}|\mathpzc{S}_{k,0})}{p_G(I_0=\{i\}|\mathpzc{S}_{k,0})} \geq \frac{p_G(I_0\neq\{k\}|\mathpzc{S}_{k,0})}{p_G(I_0=\{k\}|\mathpzc{S}_{k,0})}.
\end{equation}
For any event $\mathpzc{S}_{k,0} \subseteq \mathbb{S}$ such that $k \notin \mathpzc{Q}$, we have
\begin{equation}
\begin{split}
&\frac{p_G(I_0\neq\{i\}|\mathpzc{S}_{k,0})}{p_G(I_0=\{i\}|\mathpzc{S}_{k,0})}\\
&=\frac{\sum_{j \neq i \in \mathpzc{Q}}p_G^{(j)}(\mathpzc{S}_{k,0})}{p_G^{(i)}(\mathpzc{S}_{k,0})}\\
&=\frac{\sum_{j \neq i \in \mathpzc{Q}}\alpha(I-(1-\alpha)\hat{A}_{k})^{-1}[j,k]p_G^{(k)}(\mathpzc{S}_{k,0})}{\alpha(I-(1-\alpha)\hat{A}_{k})^{-1}[i,k]p_G^{(k)}(\mathpzc{S}_{k,0})}\\
&=\frac{\sum_{j \neq i \in \mathpzc{Q}}\alpha(I-(1-\alpha)\hat{A}_{k})^{-1}[j,k]}{\alpha(I-(1-\alpha)\hat{A}_{k})^{-1}[i,k]}
\end{split}
\end{equation}
Therefore,
\begin{equation}
\begin{split}
c &= \min_{i\in \mathpzc{Q}, k \notin \mathpzc{Q}}\bigg\{\sum_{j \neq i\in \mathpzc{Q}}\alpha(I-(1-\alpha)\hat{A}_{i})^{-1}[j,i],\\
&~~~~~~~~~~~~~~~~~~~~~~~~\frac{\sum_{j \neq i\in \mathpzc{Q}}\alpha(I-(1-\alpha)\hat{A}_{i})^{-1}[j,k]}{\alpha(I-(1-\alpha)\hat{A}_{i})^{-1}[i,k]}\bigg\},
\end{split}
\end{equation}
which completes the proof.
\end{IEEEproof}

\section{Proof of Theorem \ref{Theorem6}}\label{Prooftheorem6}
\begin{IEEEproof}
The spreading process can be considered as a Markov Chain with each state representing the number of informed nodes $N_{inf}$ in the graph. Let $X_k\triangleq \{N_{inf}=k\}$ denote the $k$-th state of the Markov Chain. For standard gossip in the asynchronous setting and private gossip (in both the synchronous and the asynchronous settings), due to the fact that only one node in the graph is active at any time during the spreading process, the state can only move from $X_k$ to $X_{k+1}$ for all $k\in \{0,1,...,n-1\}$. Each gossip action can be considered as a Bernoulli trial (successful if a previously uninformed node is informed). Given a failure probability of $f$, the corresponding successful probability decreases by a factor of $1/(1-f)$. In this sense, the expected interstate time, which is essentially the expected time between two consecutive successful gossip actions, is amplified by a factor of $1/(1-f)$. As a result, the final expected time to reach the final state, i.e., the expected spreading time is $1/(1-f)$ times that of the perfect communication scenario. Finally, private gossip is a single random walk in both synchronous and asynchronous settings, and $C_G$ is the expected time to inform all nodes in the graph.
\end{IEEEproof}

\section{Proof of Theorem \ref{delayedgeneral}}\label{Proofdelayedgeneral}
\begin{IEEEproof}
In the synchronous setting, consider two nodes $i,j$ such that $d(i,j)=D_G$, and the event that node $i$'s activity is observed by the attacker at the moment when it starts monitoring, which is denoted as $\mathpzc{S}_{i,0}$. Considering another node $k$ such that $d(k,i)=t$, the probability that $i$ is informed at round $t$ is
\begin{equation}
p_G^{(k)}(i\in I_t)\geq\prod\limits_{\substack{m\in p_{k\rightarrow i}\\ p_{k\rightarrow i}:L(p_{k\rightarrow i})=t}}\frac{1}{d_m}\geq (\frac{1}{d_{max}})^t,
\end{equation}
where $p_{k\rightarrow i}$ is a path from node $k$ to node $i$ and $L(p_{k\rightarrow i})$ is the length of this path. Then $p_G^{(k)}(\mathpzc{S}_{i,0})\geq (\frac{1}{d_{max}})^t\alpha$. It is clear that $p_G^{(j)}(\mathpzc{S}_{i,0})=0$ since it takes at least $D_G> t$ rounds for the information to be delivered to node $i$ from node $j$. Therefore, by Lemma 1, $\delta \geq (\frac{1}{d_{max}})^t\alpha$.

In the asynchronous setting, again, consider two nodes $i,j$ such $d(i,j)=D_G$, and let $\mathpzc{S}_{i,0}$ denote the event that node $i$'s activity is the first one observed by the attacker. If $j$ is the source node, denote the set of informed and active nodes after $t$ steps of communications as $INA_t(j)$. From this set, find the node $k\in INA_t(j)$ that has the shortest path to node $i$. Clearly, it requires at least $d(k,i)$ ($\geq (D_G-t)$) steps for the information to reach node $i$ from any node in $INA_t(j)$. Consider $O_{INA_t\rightarrow i}$ as the event that no communication is observed by the attacker during the process that the information flows from $INA_t(j)$ to node $i$. Then,
\begin{equation}
\begin{aligned}
p_G^{(j)}(\mathpzc{S}_{i,0})&=p_G^{(j)}(\mathpzc{S}_{i,0}\bigcap O_{INA_t\rightarrow i})\\
&\leq p_G^{(j)}(O_{INA_t\rightarrow i}) \leq (1-\alpha)^{d(k,i)}\leq (1-\alpha)^{D_G-t}.
\end{aligned}
\end{equation}
Also, considering another node $l$ such that $d(l,i)=t$, the probability that node $i$ is informed at the $t$th step from the beginning of information spreading is
\begin{equation}
p_G^{(l)}(i\in I_t)\geq(\prod\limits_{\substack{m\in p_{l\rightarrow i}\\ p_{l\rightarrow i}:L(p_{l\rightarrow i})=t}}\frac{1}{d_m})\frac{1}{t!} \geq \frac{1}{d_{max}^t t!},
\end{equation}
where $\frac{1}{t!}$ is the probability that all nodes in a path $p_{l\rightarrow i}$ are activated (whose clocks tick) in a fixed order so that the information reaches node $i$ after $t$ steps from node $l$. Finally, the probability that node $i$ is activated and its gossip action is observed by the attacker is $\frac{\alpha}{t+1}$. Therefore, $p_G^{(l)}(\mathpzc{S}_{i,0})\geq \frac{\alpha}{d_{max}^t (t+1)!}$. By Lemma 1 and the same logic as Eq. (\ref{EqP}), we have Eq. (\ref{LEq}).
\end{IEEEproof}

\section{Proof of Theorem \ref{delayedepsilon}}\label{Proofdelayedepsilon}
\begin{IEEEproof}
Let $p_{G,t}^{(i)}(\mathpzc{S}_{i,0})$ denote the probability of the attacker observing $\mathpzc{S}_{i,0}$ given source node $i$ in the scenario where the attacker starts monitoring the information spreading process $t$ steps of gossip communications after it begins. In this sense, $p_{G,0}^{(j)}(\mathpzc{S}_{i,0})$ is the probability of the event $\mathpzc{S}_{i,0}$ if the source is node $j$ and the attacker starts monitoring in the beginning of the information spreading process. In the following, the relationship between $p_{G,t}^{(j)}(\mathpzc{S}_{i,0})$ and $p_{G,0}^{(j)}(\mathpzc{S}_{i,0})$ is derived, after which similar analysis to the proof of Theorem \ref{dpprivategossip} can be applied to obtain the differential privacy level $\epsilon$. Particularly, for any $j \in V$, we have
\begin{equation}
\begin{split}
&p^{(j)}_{G,t}(\mathpzc{S}_{i,0}) \\
&=\sum_{k \in V}A^{t}[j,k]p_{G,0}^{(k)}(\mathpzc{S}_{i,0})\\
&=A^{t}[j,i]p_{G,0}^{(i)}(\mathpzc{S}_{i,0}) + \sum_{k \neq i \in V}A^{t}[j,k]p_{G,0}^{(k)}(\mathpzc{S}_{i,0})\\
&=A^{t}[j,i]p_{G,0}^{(i)}(\mathpzc{S}_{i,0}) \\
&+ \sum_{k \neq i \in V}A^{t}[j,k]\alpha (I-(1-\alpha)\hat{A}_{i})^{-1}[k,i]p_{G,0}^{(i)}(\mathpzc{S}_{i,0}).
\end{split}
\end{equation}
Therefore,
\begin{equation}
\small
\begin{split}
&e^{\epsilon} = \max_{i \in V, j\neq z \in V}\frac{p_{G,t}^{(j)}(\mathpzc{S}_{i,0})}{p_{G,t}^{(z)}(\mathpzc{S}_{i,0})}\\
&=\max_{i\in V, j \neq z \in V}\frac{A^{t}[j,i]+\sum_{k\neq i \in V}A^{t}[j,k]\alpha (I-(1-\alpha)\hat{A}_{i})^{-1}[k,i]}{A^{t}[z,i]+\sum_{k\neq i \in V}A^{t}[z,k]\alpha (I-(1-\alpha)\hat{A}_{i})^{-1}[k,i]}.
\end{split}
\end{equation}

Regarding the prediction uncertainty, recall that for any $j \in V$ and $\mathpzc{S}_{z,0} \subseteq \mathbb{S}$,
\begin{equation}
\begin{split}
p^{(j)}_{G,t}(\mathpzc{S}_{z,0})&=A^{t}[j,z]p_{G,0}^{(z)}(\mathpzc{S}_{z,0}) \\
&+ \sum_{k \neq z \in V}A^{t}[j,k]\alpha (I-(1-\alpha)\hat{A}_{z})^{-1}[k,z]p_{G,0}^{(z)}(\mathpzc{S}_{z,0})
\end{split}
\end{equation}
Then, for any $i, z \in V$, we have
\begin{equation}
\small
\begin{split}
&\frac{p_{G}(I_{0}\neq\{i\}|\mathpzc{S}_{z,0})}{p_{G}(I_{0} = \{i\}|\mathpzc{S}_{z,0})} \\
& = \frac{\sum_{j\neq i \in V}p_{G,t}^{(j)}(\mathpzc{S}_{z,0})}{p_{G,t}^{(i)}(\mathpzc{S}_{z,0})}\\
&=\frac{\sum_{j\neq i \in V}[A^{t}[j,z]+\sum_{k\neq z \in V}A^{t}[j,k]\alpha (I-(1-\alpha)\hat{A}_{z})^{-1}[k,z]]}{A^{t}[i,z]+\sum_{k\neq z \in V}A^{t}[i,k]\alpha (I-(1-\alpha)\hat{A}_{z})^{-1}[k,z]},
\end{split}
\end{equation}
which completes the proof.
\end{IEEEproof}

\end{document}